\documentclass[sn-basic]{sn-jnl}

\usepackage{amsmath}
\usepackage{hyperref}

\begin{document}
\title[Solar Cycle Observations]{Solar Cycle Observations}

\author[1]{\fnm{Aimee} \sur{Norton}}\email{aanorton@stanford.edu}
\author*[2]{\fnm{Rachel} \sur{Howe}}\email{r.howe@bham.ac.uk}
\author[3]{\fnm{Lisa} \sur{Upton}}\email{lisa.upton@swri.org}
\author[4]{\fnm{Ilya} \sur{Usoskin}}\email{ilya.usoskin@oulu.fi}


\affil[1]{\orgdiv{HEPL Solar Physics}, \orgname{Stanford University}, \state{CA},\postcode{94305-4085},  \country{USA}}
\affil*[2]{\orgdiv{School of Physics and Astronomy}, \orgname{University of Birmingham}, \orgaddress{\street{Edgbaston}, \city{Birmingham}, \postcode{B15 2TT}, \country{UK}}}
\affil[3]{\orgname{Southwest Research Institute}, \city{Boulder}, \state{CO}, \postcode{80302}, \country{USA}} 
\affil[4]{\orgdiv{Sodankyl\"a Geophysical Observatory and Space Physics and Astronomy Research Unit}, \orgname{University of Oulu}, \postcode{90014}, \country{Finland}}

\abstract{We describe the defining observations of the solar cycle that provide constraints for the dynamo processes operating within the Sun. Specifically, we report on the following topics: historical sunspot numbers and revisions; active region (AR) flux ranges and lifetimes; bipolar magnetic region tilt angles; Hale and Joy's law; the impact of rogue ARs on cycle progression and the amplitude of the following cycle; the spatio-temporal emergence of ARs that creates the butterfly diagram; polar fields; large-scale flows including zonal, meridional, and AR in-flows; short-term cycle variability; and helioseismic results including mode parameter changes.}



\maketitle
\section{Introduction}
\label{intro}

In the mid 1800s, \citet{schwabe:1844} discovered the solar cycle by observing that sunspot numbers rise and fall over the course of roughly 11 years. This discovery likely inspired Wolfe to take daily observations of the Sun, thus beginning the crucial, historical recording of the sunspot number. See Section~\ref{SN} for the history, and recent revision, of the sunspot numbers. Studies of individual ARs and their properties began with sunspot drawings and daily observations, but AR research was revolutionized by \citet{hale:1908} who demonstrated their magnetic nature, as discussed in Section~\ref{AR}.

\citet{Larmor:1919} proposed that the solar magnetic fields observed by Hale and his colleagues were generated through the process of electromagnetic induction in the electrically conducting solar plasma. This idea became known as dynamo theory and its evolution as applied to the Sun and stars is discussed extensively in \citet{charbonneau:2023}.

In addition to records of sunspot numbers, the area and position of sunspots has been recorded since 1874, beginning at the Royal Observatory, Greenwich and later continuing via the National Oceanic and Atmospheric Administration (NOAA). The distribution of sunspot area as a function of latitude and time revealed that two sunspot bands existed on either side of the heliographic equator and that these bands moved equatorward during the course of the solar cycle.This pattern is known as the butterfly diagram and is discussed in Section~\ref{BD}.  One criteria of success for any dynamo model is its ability to reproduce the features of the sunspot bands, including the observed equatorward migration. 

While sunspots are distinctive, visible features containing strong magnetic fields, sunspots account for less than 1\,\% of the solar surface area even at solar cycle maximum. Determining the larger, global-scale magnetic structure of the Sun required measurements of weaker, more spatially distributed fields \citep{Stenflo:1970}. There is a large-scale dipole field that dominates at cycle minimum. The amplitude of this dipole is a reliable precursor for the next cycle amplitude, which is best-studied through the accumulation of small-scale flux at the poles, as discussed in Section~\ref{PF}.

The magnetic nature of the solar cycle is only a part of the story. The behavior of large-scale flows informs us of the variations associated with the dynamo.  These plasma motions include differential rotation (radial and latitudinal), torsional oscillations, meridional flow and AR in-flows, as discussed in Section~\ref{Flows}. Finally, short-term cycle variations and helioseismic mode parameter changes are mentioned in Sections~\ref{Short-term} and \ref{Mode}, respectively. 

This review paper is a result of an International Space Science Institute Workshop titled ``Solar and Stellar Dynamos: A New Era" held in June, 2022.
Discussions in this review paper are intentionally brief and may not be comprehensive, since the purpose is to introduce the observations that have inspired the up-to-date research summarized in the other papers that are part of this series, \textit{Space Science Reviews:219}.

\section{Sunspot Number}\label{SN}

The sunspot number (SN) is a synthetic (not physical) quantitative index of solar activity, which is historically widely used because of its simplicity and long (more than 400 years) available dataset. 
The SN is not equal to the \textit{number of sunspots} (denoted as $s$ below) but includes also the weighted number of sunspot groups $g$, using the formula introduced by Rudolf Wolf in the middle of the 19th century:
\begin{equation}
SN = k\cdot (10\cdot g + s),
\label{Eq:WSN}    
\end{equation}
where $k$ is a scaling factor reducing the data quality (related, e.g., to the quality of the instrumentation used) of individual observers to that of the reference one (usually Rudolf Wolf or Alfred Wolfer are considered as the reference observers).
A single spot on the Sun ($s=1$) is counted as a single sunspot group leading thus to SN=11.
The classical Wolf's method uses observations of only one, so-called primary observer for each day.
If the primary observer's data was not available for a day, secondary, tertiary, etc. observers were used, but always only one per day \citep[see][]{waldmeier61}. 
This makes the SN series easy to calculate but leaves no way to verify it nor to estimate its uncertainty.
This forms the so-called \textit{Wolf} or Z\"urich SN series (WSN or $Z$). 
The WSN was continuously produced by Z\"urich Observatory using roughly nearly the same, reproducible, techniques.
The main shortcoming of the WSN is that it is not transparent and cannot be presently revisited, corrected or verified, since only the final product has been published while raw data were hand-written in log books.
These old log books are being digitized now making it potentially possible to revise the WSN in the future \citep{friedli20}.

The production of the SN series was was ceased in the 1980s in Z\"urich and smoothly transmitted to Brussels (Sunspot Index and Long-term Solar Observations project, SILSO -- \url{http://sidc.be/silso}), where it is continued in the form of the International Sunspot Number, ISN \citep{clette07}.
SILSO continues using the same formula (Eq.~\ref{Eq:WSN}) for ISN but changed the methodology so that not only a single primary observer's data, but a weighted sum of all available data are used for each day.
\begin{figure}[t]
    \centerline{\includegraphics[width=0.99\textwidth]{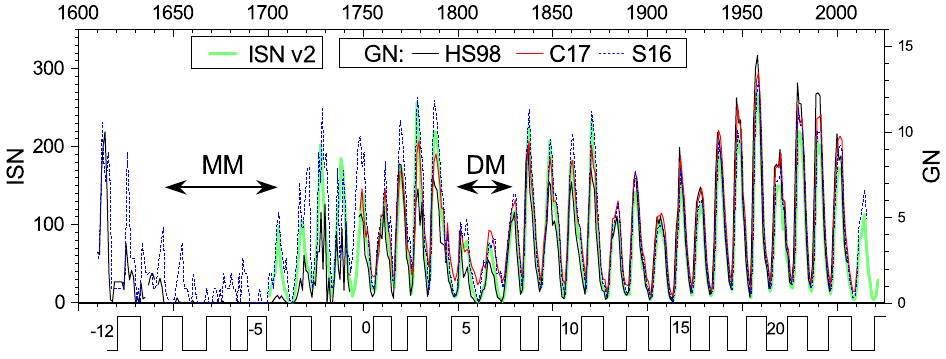}}
    \caption{Annual sunspot activity for the last centuries according to different recent reconstructions:
    International sunspot number (ISN) series version 2 (green ISN v2 curve, left axis) obtained from SILSO;
    Sunspot group number (GN, right axis), according to HS98 -- \cite{hoyt98a}; C17 -- \cite{chatzistergos17}; S16 -- \cite{svalgaard16}.
    Standard (Z\"urich) cycle numbering is shown between the panels.
    Approximate dates of the Maunder minimum (MM) and Dalton minimum (DM) are labeled within the figure.
    Modified after \citet{usoskin_LR_23}.}
    \label{Fig:SA}
\end{figure}

For more than a century, the WSN was the ``gold standard" in solar studies, but then several problems were identified, including difficulties maintaining consistency with new data. 
\cite{hoyt98a} developed a new index, called \textit{Group sunspot number}, GSN (or GN), based on a weighted sum of the number of sunspot groups reported for each day by all possible observers. 
They neglected the number of individual spots as less reliably detected.
\cite{hoyt98a} added a lot of new data not known to R. Wolf and his successors and, most importantly, published the entire database of raw data, making it possible to assess the uncertainties and add or revise the data if needed. 
This also allows evaluation of the related uncertainties.
The database of individual historical observations is continuously updated at \url{http://haso.unex.es/?q=content/data} \citep{vaquero16}.

After a careful study, several issues have been found in the WSN/ISN dataset, such as discontinuous transitions between different observers or changed methodology \citep[e.g.,][]{leussu13,clette14}.
These obvious discontinuities have been ad-hoc corrected in the revised ISN series, which have been also normalized to A. Wolfer as the reference observer -- the latter leads to a scaling factor of 1.667 with respect to the classical WSN.
This forms the ISN version 2 dataset which is considered as a current version \citep{clette16}, as shown by the green curve in Figure~\ref{Fig:SA}.
A new revision of the ISN, version 3, is pending in the near future as the first consensus dataset using the best of our present-day knowledge.

Independently of the WSN/ISN, the methodology has been revisited for the GSN series, starting from the raw-data database. 
Several new approaches have been developed in this direction. 
One was made by \cite{svalgaard16} who performed a daisy-chain `backbone' GSN composition following the classical scheme of linearly scaling individual observers between each other (blue dotted curve in Figure~\ref{Fig:SA}).
The daisy-chain approach was further improved by \cite{chatzistergos17} who accounted for non-linear relations between data from overlapping observers and composed a new GN series (red curve in Figure~\ref{Fig:SA}).
A new approach has been developed recently \citep{usoskin_ADF_16,willamo17} that uses the active-day (days with at least one sunspot observed) fraction as the metrics of the minimum size of sunspot group that could be detected by an observer due to instrumentation and seeing conditions. 

Of special interest is the level of solar activity during the Maunder minimum of 1645\,--\,1715 \citep{eddy76}: while the present paradigm is that it was nearly sunspot free \citep[e.g.,][]{usoskin_MM_15,carrasco21}, some estimates predict low but still significant sunspot activity \citep{svalgaard16,zolotova15}. 
However, a consistent analysis of the multitude of other data, such as cosmogenic isotopes, auroral records, solar eclipse observations, confirms the very low level of solar activity during the Maunder minimum \citep[e.g.,][]{usoskin_MM_15,asvestari_MNRAS_17,carrasco21,hayakawa_corona_21}, implying particularly that the reconstruction by \cite{svalgaard16} is too high in the 18th century.

Thus, at present there is a zoo of SN and GN reconstructions, as shown in Figure~\ref{Fig:SA}.
Generally, they are all consistent after about 1870 but somewhat disagree for the period between 1749\,--\,1870, with the difference being indicative of the systematic uncertainties.
The GSN series by \citet{hoyt98a} and by \citet{svalgaard16} can be considered as conservative lower and upper bounds, respectively, while other models lie between them.
A consensus-based SN reconstruction is presently not available but it is under consideration by the research community.

\section{Active Regions (ARs)}\label{AR}
An AR is identified as a dark feature observed in the solar photosphere in white light observations. ARs contain strong, bipolar magnetic fields and are associated with sunspots. In this section, we discuss observational aspects of ARs that contribute to our understanding of the solar cycle.  For a review of the origins of ARs and their emergence process, see \citet{weber:2023}.

\subsection{Hale's Law}
\begin{figure}[t]
 \centerline{
\includegraphics[width=0.9\textwidth]{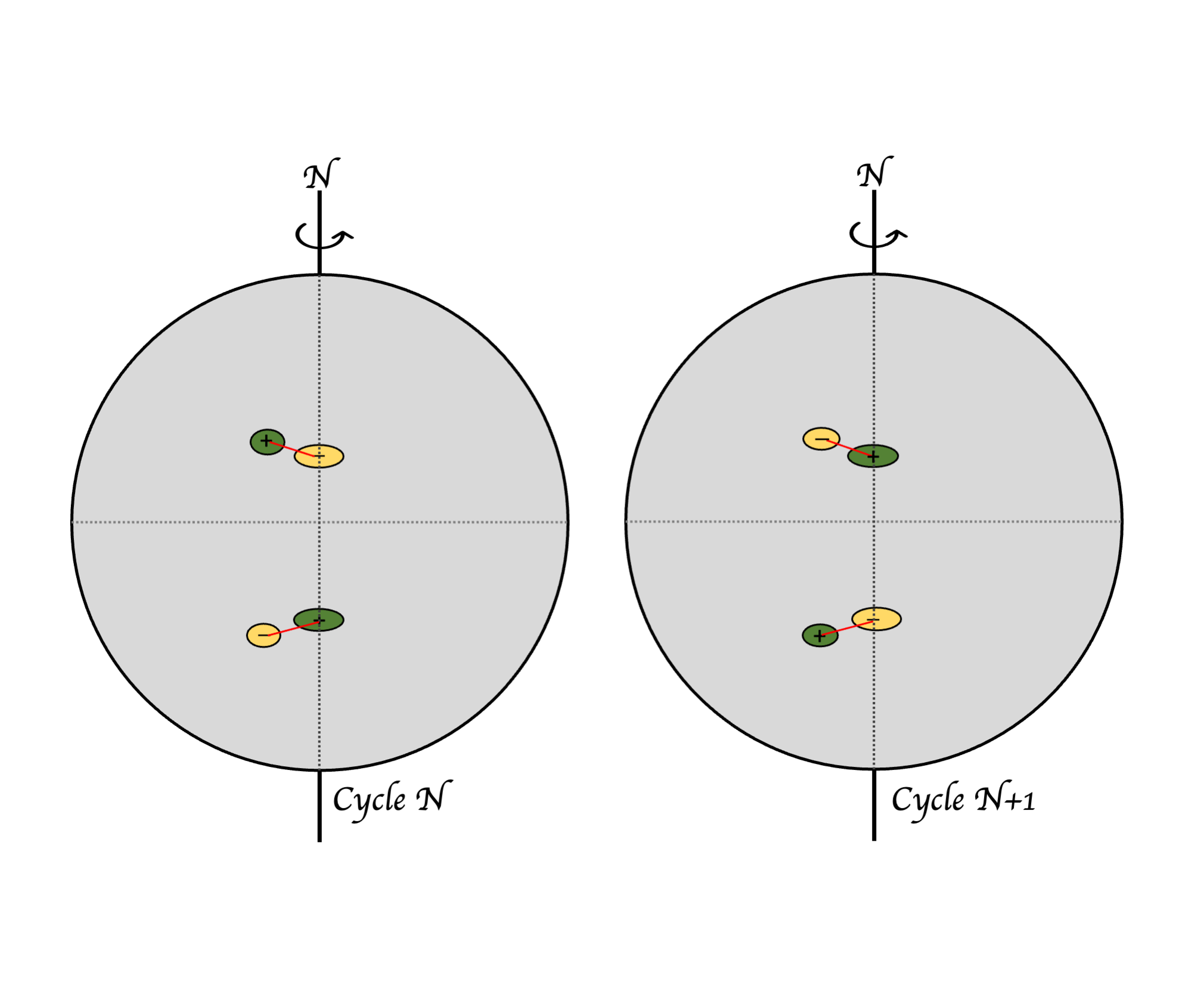}}
\centerline{
\includegraphics[width=0.9\textwidth]{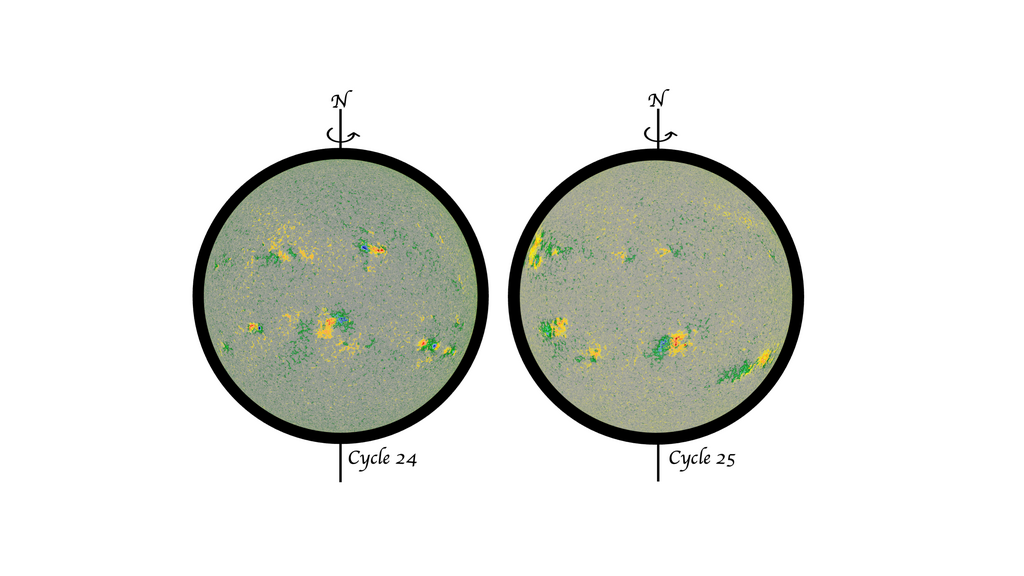}}
\caption{(Top) Hale's law describes how bipolar magnetic regions in one hemisphere tend to have the same leading magnetic polarity while those in the other hemisphere have the opposite leading polarity. This leading polarity switches from one Solar Cycle to the next, i.e., cycle N (N+1) shows the expected polarity for Cycle 24 (25). The average tilt angles between the magnetic polarities, depicted by red lines, increase with increasing latitudes. (Bottom) HMI magnetograms from Cycles 24 (2012.04.21) and 25 (2022.05.15) show the manifestation of Hale's and Joy's law on any given day with orange-red (green-blue) colors identifying the location of negative (positive) polarity. }
\label{fig:hale}
\end{figure}

\citet{hale:1908} realized that magnetic fields were the cause of sunspots after observing the Zeeman splitting of a spectral line from the light originating in a sunspot. He also noted that sunspots appeared in pairs of positive and negative magnetic polarity and that the leading polarity (with respect to rotation) in each hemisphere changes from one sunspot cycle to the next.  This is known as Hale's polarity law, see Figure~\ref{fig:hale}. While Hale's law is straightforward, it has profound implications for the solar dynamo. It implies that the large-scale organization of the magnetic field in the interior is mostly toroidal (East-West) in orientation and oppositely directed on either side of the equator. ARs adhere to Hale's law $\sim$92-95\,\% of the time \citep{wang:1989, mcclintock:2014, Li:2018, munoz-jaramillo:2021}. 

\subsection{Flux Ranges and Lifetimes}
ARs are part of a spectrum of magnetic bipoles that emerge into the photosphere and have a smooth, distribution function in regards to size and total absolute flux values ranging from 10$^{18}$ to 10$^{23}$ Mx. Ephemeral regions are smaller, short-lived regions with flux less than 1 $\times$ 10$^{20}$ Mx and have a lifetime of hours, i.e., shorter than a day \citep{hagenaar:2003}. Small regions appear as pores with flux in the range of 1 $\times$ 10$^{20}$ to 5 $\times 10^{21}$ Mx \citep{vandriel:2015}. Larger sunspots develop well-defined penumbra and have $5 \times 10^{21}$ to several $\times 10^{23}$ Mx. They live on the order of several weeks to several months. Typically, the flux emergence period is 15 – 30\,\% of the total lifetime \citep{vandriel:2015}, with most ARs fully emerged within 3\,--\,5 days \citep{harvey:1993, norton:2017} and an average emergence time of $\sim$2 days \citep{weber:2023}. After flux emergence, there is a plateau of stability before the flux begins to decay. 

\subsection{Tilt Angles}

\begin{figure}[b]
\centerline{\includegraphics[trim=-0.24in 0.0in 0.02in 0.0in,clip,width=0.700\textwidth]{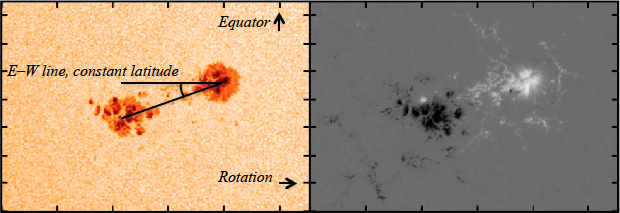}}
\centerline{\includegraphics[trim=0.00in 0.00in -0.85in 0.0in,clip,width=0.80\textwidth]{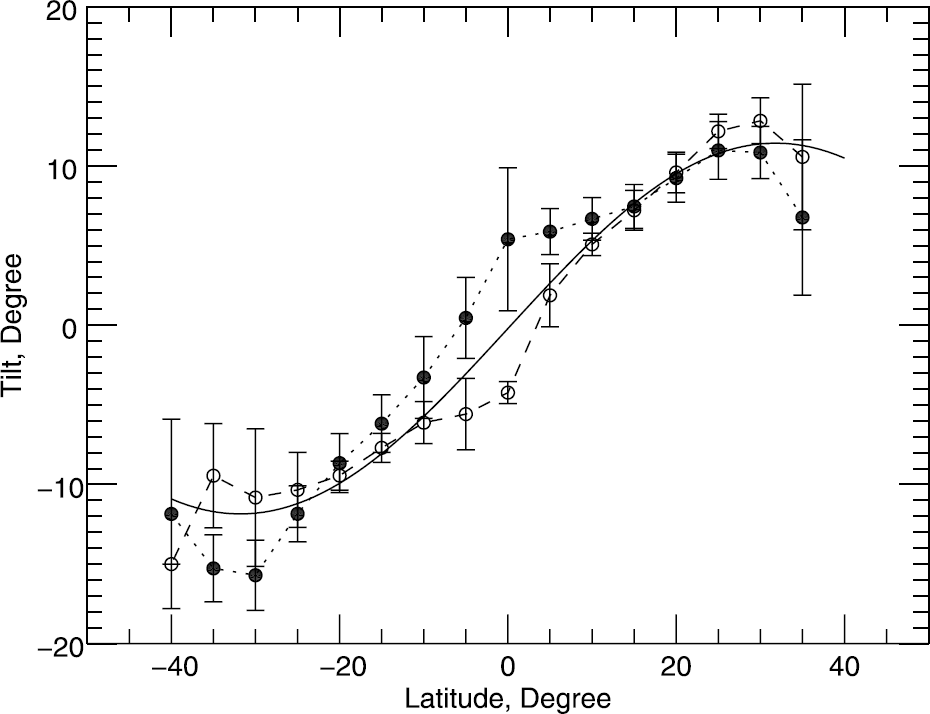}}
\caption{The majority of ARs emerge in nearly an E-W orientation with a tendency for the leading spot (with respect to rotation) to be closer to the equator. A southern hemisphere AR from Cycle 24 is shown (top) as observed in SDO/HMI intensity and magnetogram. (Bottom) The binned, mean tilt of magnetic bipoles as a function of latitude from Cycles 15$-$24 from Mt. Wilson Observatory daily sunspot drawings with magnetic polarity indicated. RMS of mean tilt shown as error bars with even (open circles, dashed line) and odd cycles (filled circles, dotted line) as well as a fit to all data (solid line) of the form $\gamma=0.2\sin(2.8\theta$) where $\gamma$ is the tilt angle and $\theta$ is latitude. This form of Joy's law captures the inflection point in tilt values at mid-latitudes. Lower image reproduced with permission from  \citet{tlatova:2018}, copyright from Springer Nature.}
\label{fig:joy}
\end{figure}

Bipolar sunspot pairs are, on average, oriented so that the leading sunspot is closer to the equator than the following sunspot, see Figure~\ref{fig:joy} for an example, and also the red lines connecting the bipolar sunspot pairs in Figure~\ref{fig:hale}, with the angle being a measure of the orientation of the bipolar magnetic region's axis with respect to the East--West direction. On average, the tilt angles increase with latitude, and this trend was named ``Joy's Law" by \citet{zirin:1988}, see Figure~\ref{fig:joy}. Tilt angles are crucial in flux-transport dynamo models where it plays a role in the formation and evolution of polar fields \citep[see, e.g.,][]{wang:1991, dikpati:1999}. Tilts serve as an observable feature of the conversion of toroidal magnetic field into poloidal, i.e., the $\alpha$-effect, and the reversal of axial dipole between cycles \citep{cameron:2018}.
There are two dominant, physical explanations for the origin of Joy's law. First, as proposed by \citet{babcock:1961}, the tilt angle observed in the photosphere reflects the directional components of the global magnetic field at depth and is a direct consequence of the ``winding up" of the poloidal field in the solar interior. Second, \citet{wang:1991} propose that Joy's law is a result of the Coriolis force acting on flows within the flux tube as it rises through the convection zone. 

Joy's law is a statistical law and only becomes obvious after much averaging. A study by \citet{wang:1989} with over 2500 bipolar magnetic regions reported that 16.6\,\% had no measurable tilts, 19\,\% were anti-Joy, 4.4\,\% were anti-Hale.  That is 39.9\,\% of regions that were not obeying Joy's law. The data are so noisy that Joy's law cannot be recovered for Cycle 17 (Cycle 19) in the northern (southern) hemisphere, respectively \citep{mcclintock:2013}  The scatter is thought to have a physical origin, the buffeting of flux tubes by convective motions \citep{fisher:1995, weber:2011}. In addition to the high scatter of tilt angles, the expansion of the sunspot group along its major axis is observed with the possibility of differential rotation acting on the poleward and equatorward spots accordingly \citep{gilman:1986, schunker:2020}.

Simulations of thin flux tubes rising through the convection zone with the Coriolis force acting on flows within the flux tube have been able to recreate both Joy's law and its scatter \citep{d'silva:1993, fan:2009, weber:2011} with scatter increasing for flux tubes that spend a longer time rising. Results from 3D dynamo models show some promise of producing bipolar magnetic regions that adhere to Hale and Joy's law \citep{nelson:2013}; there is certainly no consensus as to which models most accurately represent solar conditions and recreate tilt angle distributions.

Various forms of Joy's law are reported in the literature including the following where $\gamma$ is the tilt angle and $\theta$ the latitude:

\begin{equation}
    \sin\gamma = m\cdot \sin\theta + c,
\label{Eq:JL1}
\end{equation}
\begin{equation}
    \gamma = m\cdot \theta + c,
\label{Eq:JL2}
\end{equation}
\begin{equation}
    \gamma = m\cdot \sin\theta + c,
\label{Eq:JL3}
\end{equation}
\begin{equation}
    \gamma = m\cdot \sin(k\cdot \theta), 
\label{Eq:JL4}
\end{equation}



The reported best-fit values for slope, $m$, and intercept, $c$, depend on the solar cycle, instrument, type of data (white-light or magnetogram), and sampling techniques used for the determination of the tilt angles. Note that some researchers force the fits through zero, $c=0$, while others allow a $y$-intercept that is non-zero. For a few examples, \citet{wang:1991} report Joy's law in the form of Equation \ref{Eq:JL1} with $m=0.48$ and $c=0.03$, \citet{norton:2005} use Equation \ref{Eq:JL2} with $m=0.2$ and $c=0.2$, \citet{tlatova:2018} uses Equation \ref{Eq:JL4} with $m=0.2$ and $k=2.8$. \citet{Li:2018} comprehensively reports fits to the forms of Equations \ref{Eq:JL1}\,--\,\ref{Eq:JL3}. These are only a few examples, as it is beyond the scope of this review to report an all-inclusive list of Joy's law fits.

A list of observational aspects of tilt angles are as follows: 
\begin{itemize}
\item the dependence on latitude differs between cycles and hemispheres \citep{dasi-espuig:2010, mcclintock:2013, tlatova:2018}; 
\item there is evolution as the AR emerges and decays so the time of measuring a tilt angle matters \citep{mcclintock:2016, schunker:2020}; 
\item scatter is higher during the first day of emergence \citep{schunker:2020}, 
\item the value tends to settle near the end of emergence \citep{stenflo:2012, schunker:2020}; 
\item there are conflicting reports as to whether the tilts show a dependence on magnetic flux as predicted in thin flux tube modeling \citep{fisher:1995, stenflo:2012, jiang:2014, mcclintock:2016, schunker:2020}; 
\item the scatter in the tilt values has a dependence on flux but not latitude \citep{fisher:1995}; 
\item the mean and median tilts of regions near the equator are not zero indicating that forcing a fit for Joy's law through the origin may be unphysical, 
\item an inflection point in the fit of tilts as a function of latitude occurs around $30^{\circ}$ in both hemispheres \citep{tlatova:2018}; 
\item the smallest bipoles appear to have negative tilts \citep{tlatov:2013}; 
\item and the anti-Hale regions may not simply be the tail of the distribution of tilt angles as \citet{munoz-jaramillo:2021} reports they prefer an east-west orientation and have a distribution distinct from the ARs that follow Joy's law. 
\end{itemize}

Improvements in tilt angle measurements and databases is ongoing work. Traditional
determinations of Joy's law have been based on white-light images because magnetograms only became routinely available in the mid-1960s. White-light studies yield median tilt angles that are smaller and increase less steeply with latitude (lower slopes) than those obtained from magnetic data as shown by \citet{wang:2015}, who also pointed out that a substantial fraction of tilts determined from white-light data were erroneous since they were from sunspots of the same polarity. In addition, if plage is included in the calculation, the tilt angle is usually higher. Given the errors in tilt angle determinations prior to routine magnetograms, and inconsistent methodologies (i.e., tilt angles determined including only umbra versus those determined using umbra, penumbra and plage), it is not clear if long-term trends of tilt angles using only white-light data are valid. 

An anti-correlation between area-weighted mean tilt angles (normalized by latitude) and cycle strength was shown by \citet{dasi-espuig:2010} for cycles 15-21, indicating that the surface source for the poloidal field becomes weaker for stronger cycles, potentially limiting the strength of the next cycle, and providing a feedback mechanism (``tilt-quenching") that prevents runaway solutions to the cycle amplitude.  However, \citet{mcclintock:2013} could only recover the \citet{dasi-espuig:2010} result for the Southern hemisphere, not the Northern, and the Cycle 19 outlier value dominated the fit for the Southern hemispheric data. Nevertheless, non-linear feedback mechanisms that affect average tilt angles appear effective. Surface flux-transport modeling by \citet{cameron:2012} and \citet{jiang:2010} incorporated the effect of AR inflows into surface flux transport models and found that strong cycles produce strong in-flows which result in a lower tilt angle and decreased resulting axial dipole moment. 

\subsection{Rogue Active Regions}
The progression of any solar cycle, including the polarity reversal and gradual strengthening of the polar caps responsible for the axial dipole moment, is punctuated by the appearance of unusually influential, or rogue, ARs. The term ``rogue AR" was coined by \cite{Nagy:2017} who reported that a single rogue bipolar magnetic region in their simulations was found to have a major effect on the development of subsequent solar cycles, either increasing or decreasing the amplitudes, and in extreme cases, triggering a grand minimum.  \cite{Nagy:2020} then proposed the AR Degree of Rogueness (ARDoR) quantity that is the difference between the final contribution to the axial dipole moment from an individual AR and an ideal contribution from a region at the same latitude that has an expected tilt angle prescribed by Joy's Law and separation of opposite polarity footpoints typical for an AR of similar flux. Meaning, a region is defined as rogue when its contribution to the final axial dipole moment is significantly different from an average active region emerging at the same latitude. 

\cite{Petrovay:2020a} formulated an algebraic method that consists of summing the ultimate contributions of individual ARs to the solar axial dipole moment at the end of the cycle. \cite{Nagy:2020} performed a statistical analysis of a large number of simulated activity cycles and ranked the ARs from most to least influential depending on their contribution to the final axial dipole moment. The model by \citet{lemerle:2017} used in the simulation couples a conventional surface flux transport (SFT) 2D simulation defined over a spherical surface with a 2D axisymmetric flux transport dynamo (FTD) simulation defined in a meridional plane \citep{char:2005}. In this hybrid 2$\times$2D Babcock-Leighton dynamo model, the SFT component provides the surface poloidal source term for the FTD simulation, while the FTD component provides the magnetic emergence events input to the SFT simulation \citep{char:2005,Nagy:2017}. They showed that the top 50 influential ARs of any given cycle are sufficient to reproduce the final dipole moment of that cycle. Rogue ARs have a variety of characteristics but their rogueness is commonly determined by having one or more of the following characteristics: a very large amount of flux ($\Phi>1 \times 10^{22}$ Mx); an abnormal tilt angle such as one that is anti-Joy or anti-Hale (90--180$^{\circ}$ away from Joy's law); an unusually large separation distance between the polarities ($>$ 70 Mm); or being very close to the equator. 

\section{The Butterfly Diagram}\label{BD}

 \begin{figure}[t]
 \noindent\includegraphics[width=\textwidth]{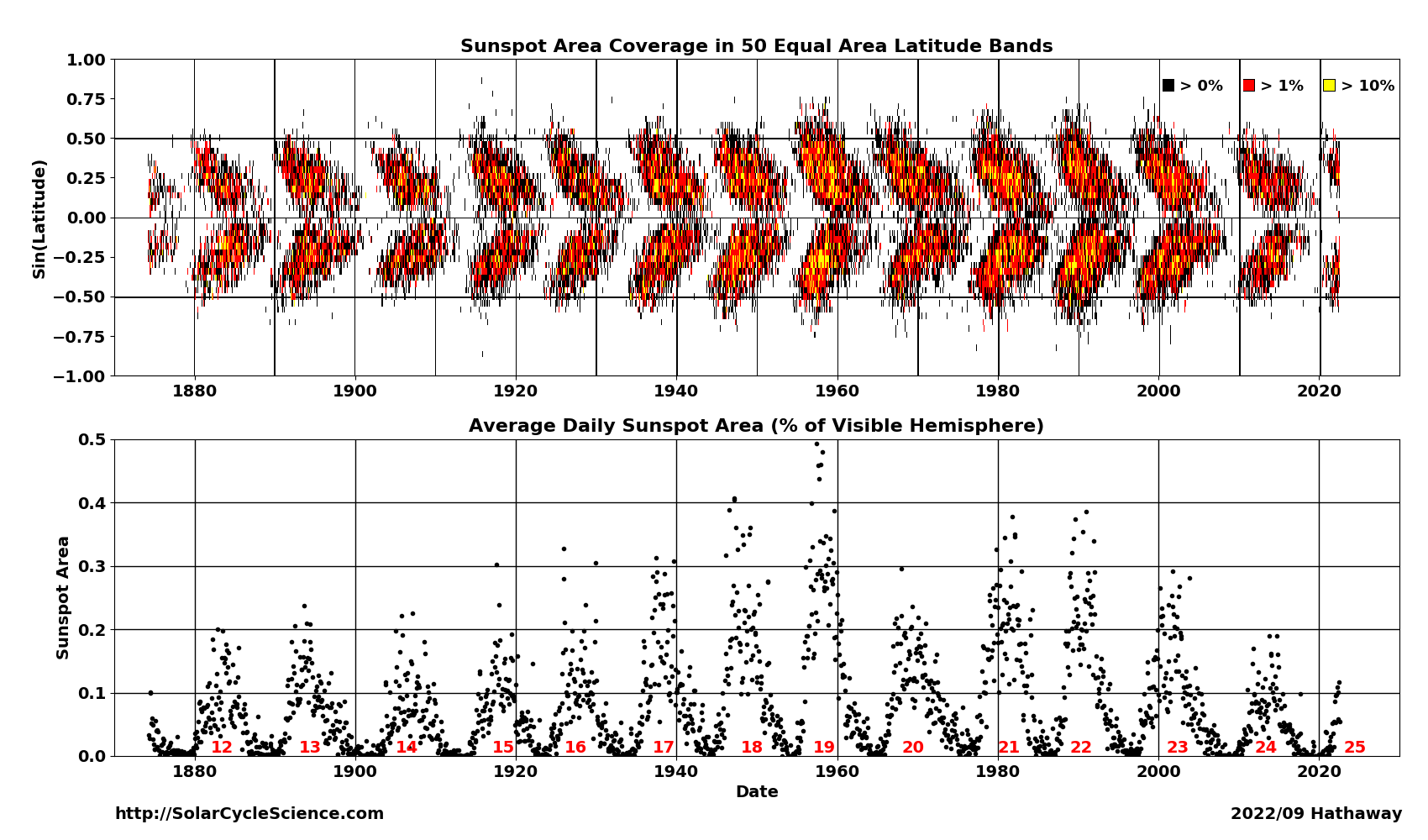}
\caption{Butterfly diagram shows sunspot area from the Royal Greenwich Observatory, color coded as a percentage of the solar disk, plotted as a function of time and sine latitude (top panel). For reference, the total sunspot area is also plotted as a function of time (lower panel). Figure courtesy D.H. Hathaway via \url{www.solarcyclescience.com}.
}
\label{fig:bfly}
\end{figure}

When the location of sunspots or ARs are plotted as a function of latitude and time, a striking pattern emerges that resembles butterfly wings. This so called ``Butterfly Diagram", first depicted by \citet{maunder:1904}, shows bands of sunspot activity in both the Northern and Southern hemispheres. Modern depictions include a third dimension, the 
fractional area of the Sun covered by sunspots, see Figure~\ref{fig:bfly}. Inspection of the butterfly diagram reveals that early in the cycle, ARs begin emerging at mid-latitudes (approximately 30 degrees) and as the cycle progresses the emergence moves closer to the equator. This equatorward progression of AR emergence is known as Spörer's Law \citep{maunder:1903}. Stronger cycles tend to begin emergence at higher latitudes than weaker cycles. The latitudinal width of the ``Butterfly wings" also changes over the course of the cycle and is proportional to the strength of the solar cycle \citep{ivanov:2011}, producing a tapering of the wings at both the start and the end of the cycle.
Typically the cycles overlap in time by about one\,--\,two years, with the new cycle beginning at mid-latitudes before the previous cycle has finished. However, this overlap is proportional to the strength of the following cycle such that the weakest cycles have little to no overlap with the previous cycle. 
Asymmetry between the northern and southern hemispheric cycle progression was noted by \citet{sporer:1894} and \citet{maunder:1904} with unequal sunspot activity persisting for several years. \citet{norton:2014} showed that the hemispheric asymmetry never had more than a 20\% difference in the amplitude of sunspot number or sunspot area. Nor was the time lag (measured by time of polar reversal, cycle maximum or minimum in each hemisphere) more than 20\,\% of the entire cycle length. In other words, the hemispheres are strongly coupled - to within $\approx$80\,\%.

\begin{figure}
 \noindent\includegraphics[trim=0.0in 0.3in 1.0in 0.0in,clip,width=1.0\textwidth]{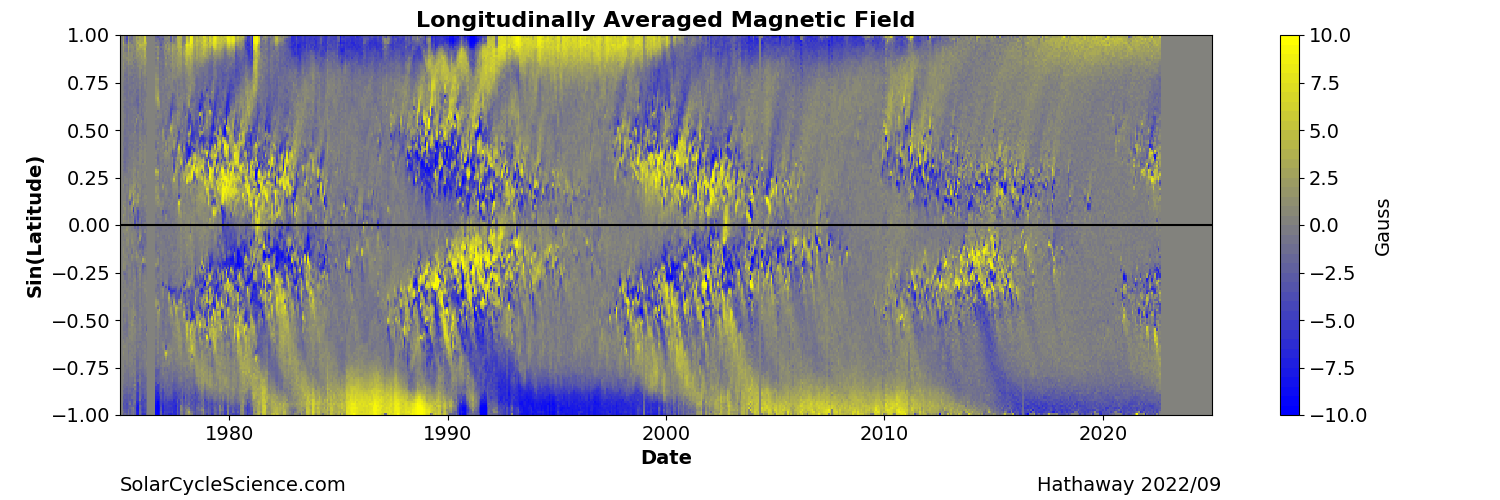}
\caption{The magnetic butterfly diagram shows the latitudinal distribution of the magnetic field as a function of time. The magnetic field is averaged over all available longitudes and over each Carrington Rotation using data from SOLIS/MDI/HMI. The color indicated the sign of the polarity, with yellow (blue) for positive (negative) radial magnetic fields. Figure courtesy D.H. Hathaway via \url{www.solarcyclescience.com}. }
\label{fig:magbfly}
\end{figure}

While the traditional Butterfly Diagram is created in statistical manner by plotting the sunspot area as a function of latitude and time, a similar plot can be created by plotting the longitudinally averaged magnetic field instead \citep{harvey:1994}. This ``Magnetic Butterfly Diagram", see Figure~\ref{fig:magbfly} reveals several other characteristics of the solar cycle. Most notable are the appearance of Joy's Law and Hale's law. Each wing displays predominantly leading polarity on the southern edge and the opposite following polarity predominantly on the northern edge (Joy's Law). The wing polarity is opposite across hemispheres and switches from one cycle to the next (Hale's law). In addition to the butterfly wings, streams of flux can be seen emerging from the wings and moving towards the poles. They are most prominent during solar cycle maximum and are dominated by the following sunspot polarity flux for that cycle (though intermittent leading-sunspot polarity streams are also present). These streams are a signature of the pole-ward meridional flow transporting residual AR flux to the polar regions. This process forms strong flux concentrations at the poles, i.e. the polar fields, that reverse polarity near the time of solar maximum.     

\section{Polar Fields}\label{PF}
\label{Sec:Polar}

During solar minimum, the Sun's magnetic field resembles that of a dipole, with opposite polarity magnetic field concentrations at the poles. This dipolar magnetic field  acts as the seed field for the solar cycle described in \citet{charbonneau:2023, cameron:2023}. 

The Sun's polar magnetic fields can be measured by averaging the magnetic field strength over the polar cap to get the flux density over each polar region or by calculating the axial dipole moment of the magnetic field configuration. The latter provides a single value for the state of the Sun's global magnetic field as a whole, while the former provides additional information about the differences between the North and South hemispheric polar magnetic field. 

 \begin{figure}[t]
 \noindent\includegraphics[width=\textwidth,trim={105 10 70 10},clip]{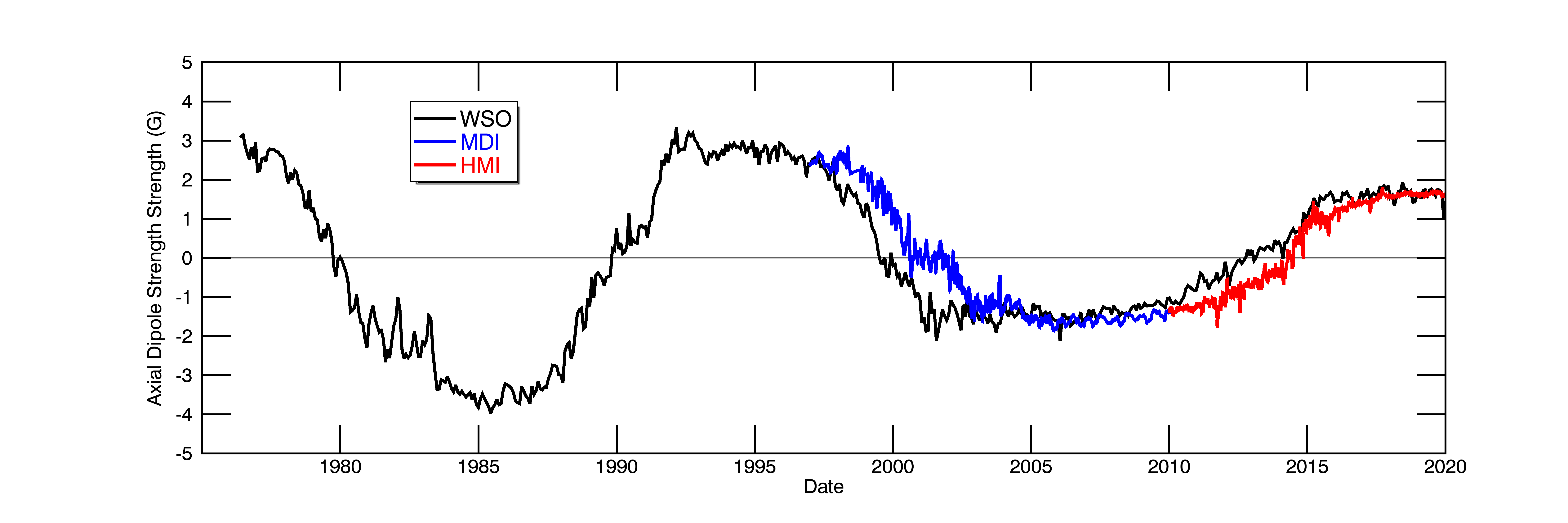}
 \noindent\includegraphics[width=\textwidth,trim={100 0 70 40},clip]{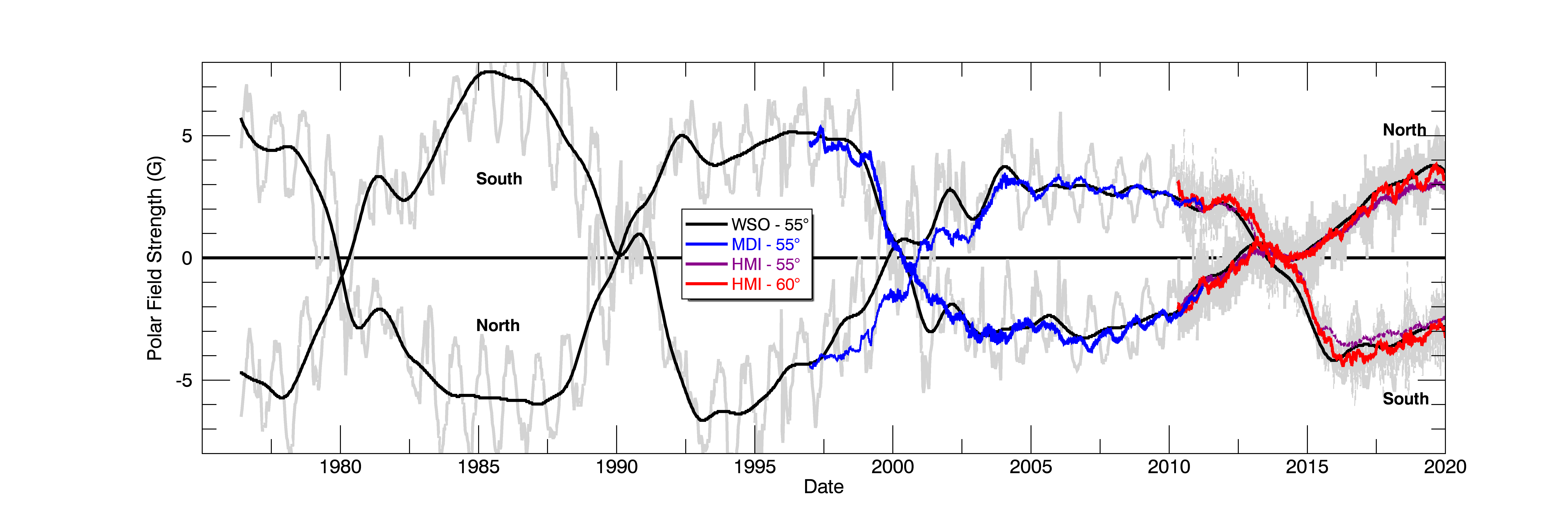}
\caption{The Sun's polar fields measured by the axial dipole moment (top) and by the average field strength over the polar caps (bottom). Data is shown from WSO in black, SOHO/MDI in blue, and SDO/HMI in purple and red. WSO and MDI averaged polar fields are measured from $55^{\circ}$ and above, while HMI is shown for both $55^{\circ}$ and $60^{\circ}$ (purple and red) and above. The polar fields are smoothed over 13 Carrington Rotations (for reference, the unsmoothed WSO and HMI measurements are shown in grey, highlighting the annual signal caused by the changing inclination of the Sun over Earth's yearly orbit). 
}
\label{fig:pfields}
\end{figure}

The Wilcox Solar Observatory (WSO) has been measuring the Sun's line-of-sight magnetic field daily since 1976 and has provided measurements of both the polar field strength and the axial dipole since that time\footnote{ \href{http://wso.stanford.edu/Polar.html}{} courtesy of J.T. Hoeksema.} \citep{svalgaard:1978,Hoeksema:1995}, and the axial dipole component is shown as a black line in Figure~\ref{fig:pfields} (top panel).  
The axial dipole moment is an integrated quantity that measures the axisymmetric component of the large-scale photospheric magnetic field. The polar field strength, bottom panel in Figure~\ref{fig:pfields}, is defined as the flux density of the magnetic field above a specific latitude. For WSO this is limited by the spatial resolution and taken to be the line of sight field strength measured in the highest-latitude pixel, which is taken to be between 55$^\circ$ and the poles (but the actual latitude range varies with the Earth's orbit). Space based missions have a better resolution and in the case of HMI\footnote{ \href{http://jsoc.stanford.edu/data/hmi/polarfield/}{} courtesy of Xudong Sun.} calculate the polar fields as the inferred radial component of the magnetic field measured at 60$^\circ$ and above \citep{sun:2015}. While the improved resolution does mitigate the projection effects, a residual annual oscillation (gray lines in the bottom panel in Figure~\ref{fig:pfields}) is evidence that there is still uncertainty in these measurements due to the poor viewing angle. While the flux density over each polar region offers insight into hemispheric asymmetries, the innate ambiguity associated with this measurement may make the axial component of the Sun's magnetic dipole a better metric for solar cycle prediction \citep{upton:2014a}.

The polar fields are out of phase with the sunspot number, with the reversal occurring near the time of solar cycle maximum. The peak in the polar field strength typically occurs at or just before solar cycle minimum. The amplitude of the Sun's polar fields (as measured by the axial dipole or by the field strength over the polar cap)  at the time of solar cycle minimum are proportional to the amplitude of the next solar cycle. Consequently, the amplitude of the polar fields at the time of cycle minimum have proven to be successful predictors of solar cycle amplitude \citep{1978Schatten_atal, 2005Svalgaard_etal, 2010Petrovay, 2013MunozJaramillo_etal, bhowmik:2023}. This can be understood in terms of the dynamo model proposed by \citet{babcock:1961} and extended by \citet{Leighton:1964}. For a detailed account of the Babcock-Leighton model, see \cite{charbonneau:2023} and \cite{cameron:2023}.

\begin{figure}[t]
    \centering
    \includegraphics[width=0.70\linewidth]{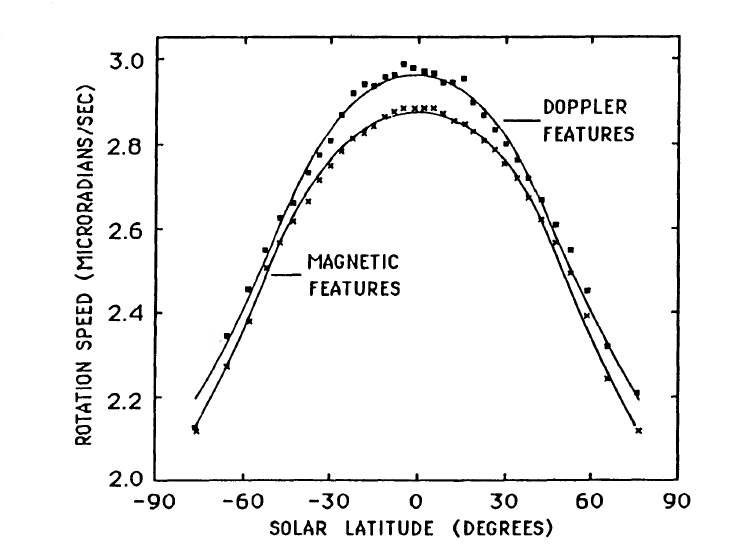}
    \includegraphics[width=0.70\linewidth]{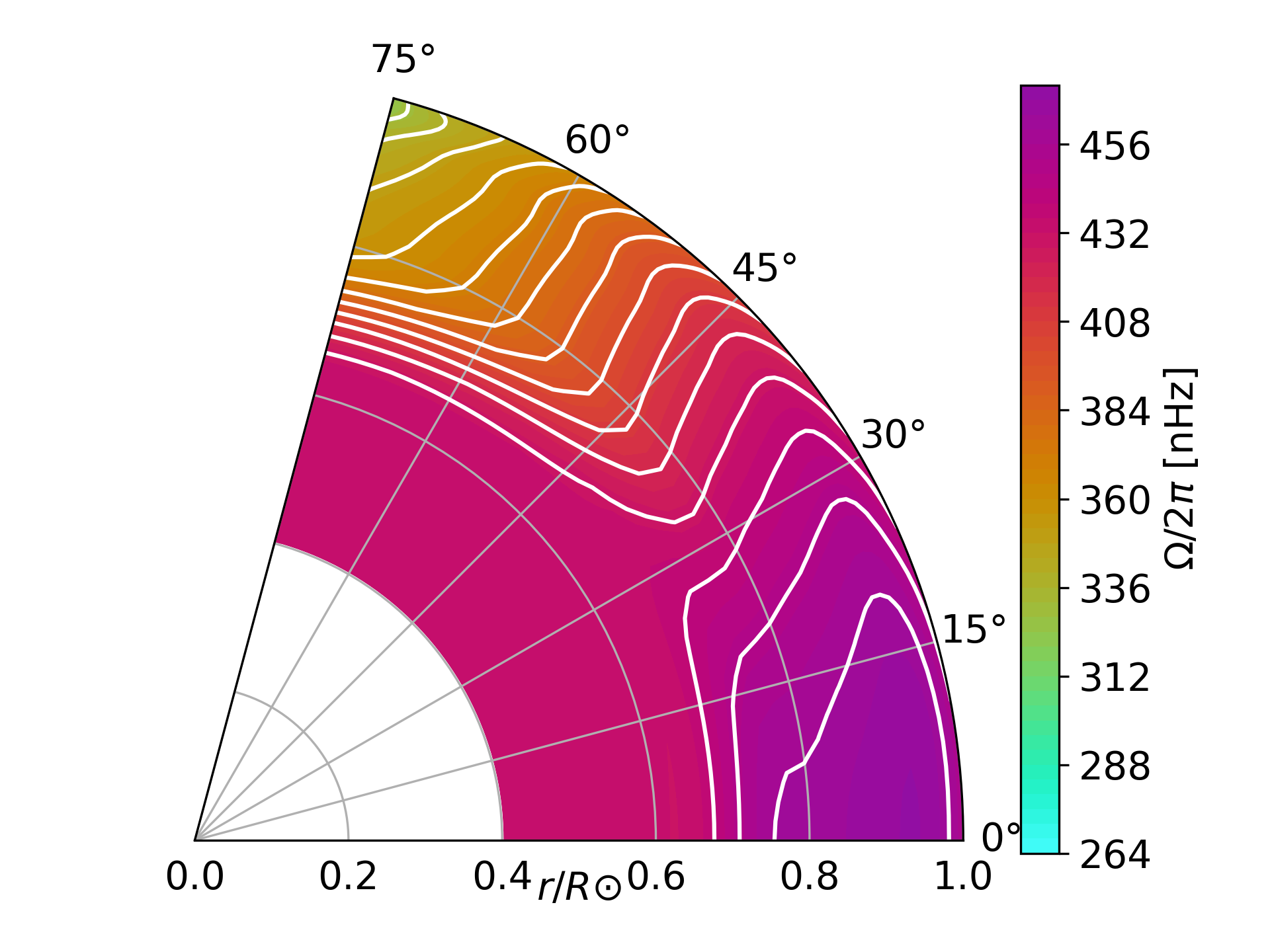}
    \caption{Top: Differential rotation measured by feature tracking \citep[from ][\copyright the AAS, reproduced by permission]{1990ApJ...351..309S}. Bottom: Solar rotation profile from 2D inversions of HMI helioseismic observations, averaged from 2010 to 2022, prepared by the authors for this review.}
    \label{fig:rotation}
\end{figure}

\section{Flows}\label{Flows}

The observed large-scale flows of the Sun --- differential rotation, torsional oscillations, meridional circulation, large-scale convection and the recently observed inertial modes --- provide a set of measurements that characterize the solar convective processes. For a detailed account of the plasma flows in the Sun, including Rossby waves and inertial modes, see \cite{hotta:2023}. Herein, we introduce the fundamental observations of these flows.  

\subsection{Solar Rotation Profile}

The mean solar rotation profile is well known. At the surface the latitudinal differential rotation can be measured by tracking features such as sunspots, revealing that the rotation rate is highest at the solar equator and decreases towards the poles (Figure~\ref{fig:rotation}, top). A comprehensive review of these measurements has been made by \citet{2000SoPh..191...47B}. Helioseismology \citep[see, for example,][]{1996Sci...272.1300T, 1998ApJ...505..390S, 2018SoPh..293...29L} has revealed the interior rotation profile (Figure~\ref{fig:rotation} bottom). It features a near-surface shear layer (sometimes abbreviated as NSSL) where the rotation rate increases with depth down to about $0.95\,{\mathrm R}_{\odot}$. Below this layer, latitudinal differential rotation persists through the bulk of the convection zone, approximately constant on radial lines although the isorotation contours tend to lie at about a 25-degree angle to the rotation axis over a wide range of latitudes \citep{2003soho...12..283G}. There is another shear layer or ``tachocline'' at the $0.71\,{\mathrm R}_{\odot}$ base of the convection zone, which is narrower in reality than it appears in most helioseismic profiles due to the finite resolution of the inversions; the consensus \citep[see Table 2 of][and references therein]{2009LRSP....6....1H} is that the thickness is around $0.05\,{\mathrm R}_{\odot}$, but at least one estimate \citep[][]{1999A&A...344..696C} puts it as low $0.01 \,{\mathrm R}_{\odot}$. Below the tachocline, in the radiative interior, there is roughly rigid rotation down to the limits of reliable measurement at around $0.2 \,{\mathrm R}_{\odot}$. \citep[e.g.][]{1998ESASP.418..685E, 2003ApJ...597L..77C}, although the former authors note that it is possible that the 
core is rotating somewhat faster than the bulk of the radiative interior). 

\begin{figure}[t]
    \centering
    \includegraphics[width=0.99\linewidth]{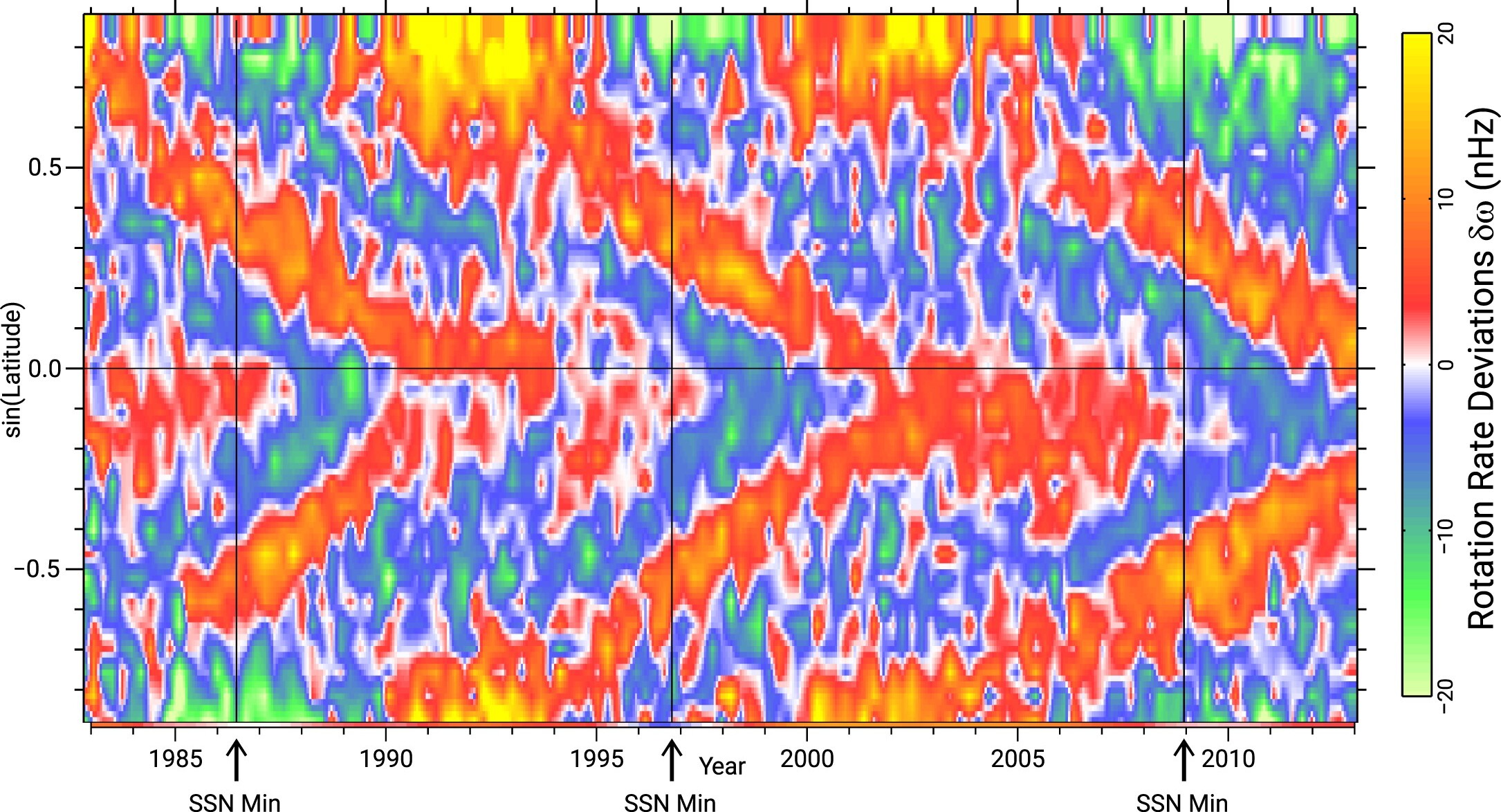}
    \caption{The torsional oscillation in Mount Wilson surface Doppler observations, adapted from \citet{Ulrich_2022} under the \href{http://creativecommons.org/licenses/by/4.0}{CC BY 4.0} license.}
    \label{fig:ulrich}
\end{figure}

\subsection{Zonal Flows: Torsional Oscillations}
\begin{figure}[t]
    \centering
    \includegraphics[trim=-0.5in 0.in 0.4in 0.2in,clip,width=\linewidth]{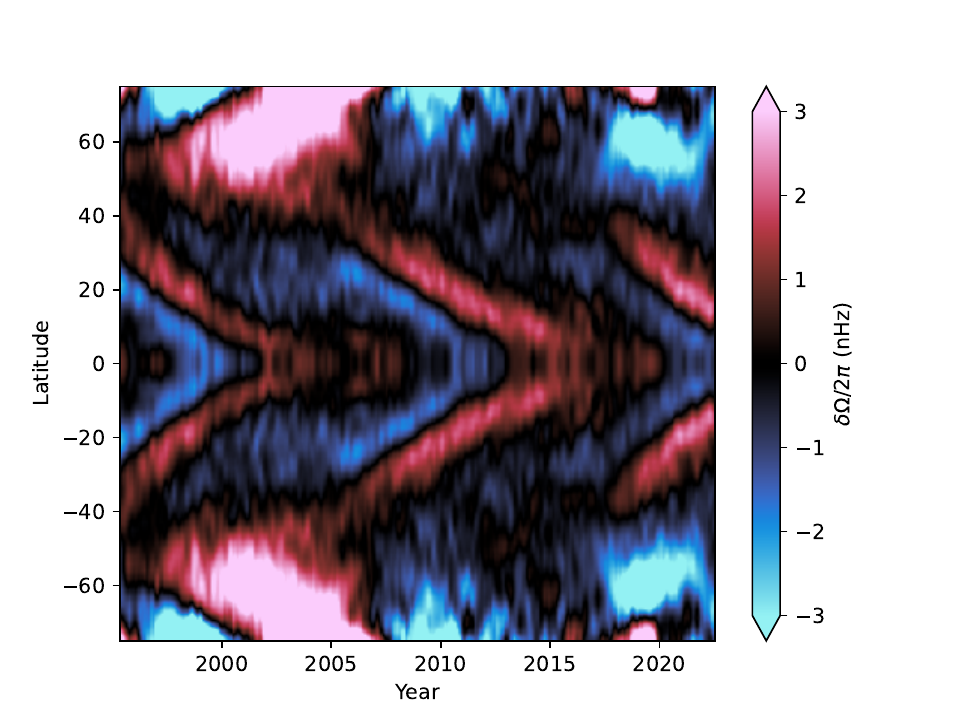}
    \caption{Zonal flow map from helioseismic inversions of GONG (1995\,--\,2022), MDI (1996\,--\,2010), and HMI (2011\,--\,2022) data, at a target depth of $0.99\,R_{\odot}$, with a temporal mean over the whole dataset subtracted at each latitude. Note that $\delta\Omega$ here is the same quantity as $\delta\omega$ in Figure~\ref{fig:ulrich}. Reproduced from \citet{Howe_2022} under the \href{http://creativecommons.org/licenses/by/4.0}{CC BY 4.0} license.}
    \label{fig:rh_zonal}
\end{figure}

The solar rotation profile is modulated by a pattern of bands of faster- and slower-than-average rotation, which can be considered respectively as prograde or retrograde flows, and which migrate in latitude in synchrony with the solar cycle (see Figures~\ref{fig:ulrich} and \ref{fig:rh_zonal}). This pattern, revealed when a temporal average is subtracted from the rotation rate at each latitude, was first observed, and dubbed the ``torsional oscillation'', by \citet{1980ApJ...239L..33H} in surface Doppler observations from the 150 ft tower at the Mount Wilson Observatory. The Mount Wilson observations continued until 2013, and Figure~\ref{fig:ulrich} from \citet{Ulrich_2022} shows the pattern over three solar cycles. The main feature is the band of faster rotation in each hemisphere that moves from mid-latitudes towards the equator between one solar minimum and the next; as pointed out by  \citet{1980ApJ...239L..33H}, the latitude of maximum flux falls close to the edge of this belt. These flows are relatively weak compared to the mean solar rotation, with amplitudes close to the surface of less than ten meters per second, or a fraction of a per cent of the equatorial rotation rate. 

The flow patterns were seen in helioseismic data in the rising phase of Solar Cycle 23 by \citet{1999ApJ...523L.181S}, and \cite{2000ApJ...533L.163H} found that the patterns penetrated at least 0.08 solar radii into the convection zone; subsequent work, for example by \citet{2002Sci...296..101V}, suggested that the variation in rotation involves most of the bulk of the convection zone. A strong band of faster flow migrating from mid-latitudes towards the poles early in Solar Cycle 23 was reported by \citet{2001ApJ...559L..67A}.

Because the mid-latitude rotation begins to speed up before significant surface activity is seen, the flow pattern towards the end of one solar cycle can give some indication of the timing of the onset of the following one, as reported by \citet{2009ApJ...701L..87H} for Cycle 24 and \citet{2018ApJ...862L...5H} for Cycle 25. In particular, the time at which the main belt of faster rotation reaches a latitude of around 25 degrees seems to coincide with solar activity becoming widespread in a new cycle. 
The strong poleward branch seen in Cycle 23 was not repeated in Cycle 24 \citep{2013ApJ...767L..20H}. This seems to be associated with small but significant deceleration at higher latitudes, possibly related to the weaker polar fields in Cycle 24 \citep{2012ApJ...750L...8R}.
Figure~\ref{fig:rh_zonal} shows the flow residuals from inversions of GONG, MDI, and HMI data, as reported by \citet{Howe_2022}. We note that the global helioseismic inversions can only show the North--South symmetric part of the flow pattern, while the surface measurements and those from local helioseismology \citep[e.g.][]{2018SoPh..293..145K, 2018ApJ...861..121L} can distinguish the two hemispheres. The relationship between the flow pattern and magnetic butterfly diagram is complex, but \citet{2018ApJ...861..121L} found that the hemispheric asymmetry of the flows is related to, and is a leading indicator of, the magnetic asymmetry; asymmetry in the flows is seen in advance of the corresponding asymmetry in the magnetic activity.

\subsection{Meridional Flows}

The solar meridional flow is the North-South motion of the plasma. At the surface this flow plays a critical role in the solar dynamo by transporting residual flux from ARs to the poles in order to generate the magnetic field to initialize the next solar cycle. This plasma flow moves from the equator to the poles in each hemisphere with an amplitude of $\sim 10-20$ m s$^{-1}$. The meridional flow is 1-2 orders of magnitude weaker than the differential rotation (relative velocities of $\sim 200-250$  m s$^{-1}$) and the convective flows (velocities of $\sim 500$ m s$^{-1}$ for supergranules and $\sim 3000$  m s$^{-1}$ for granules), making it the most challenging plasma flow to measure. The meridional flow is typically measured in the same manner as (and along with) the differential rotation (e.g., Doppler imaging, helioseismology, tracking techniques, etc.). Characterizing this flow is particularly challenging because independent measurement techniques can often give very different measurements, thought to be a consequence of the different depths sampled by each technique. For an in depth review, see \citet{hanasoge:2022} and references therein.

\begin{figure}[t]
    \centering
       
        \includegraphics[trim=0.0in .0in 0.0in .0in, clip, width=0.94\linewidth]{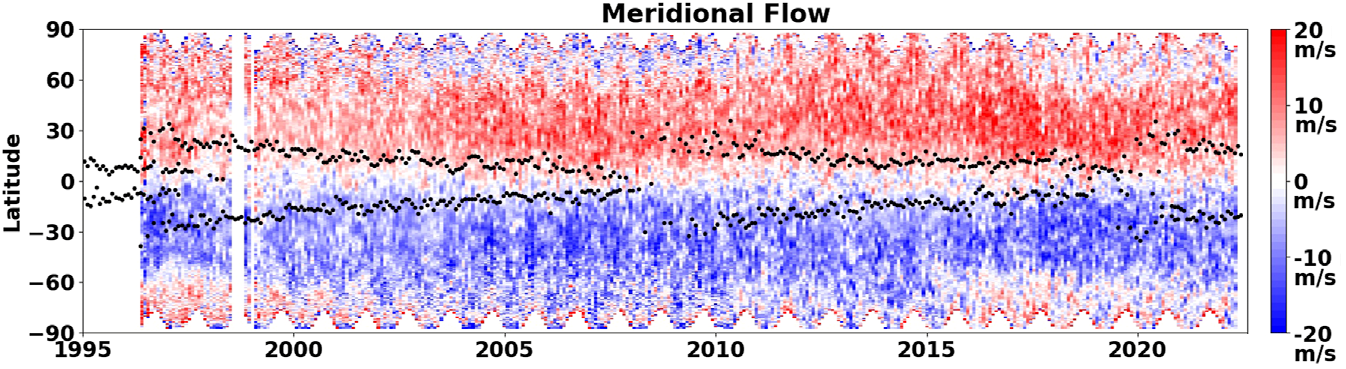}

    \caption{The evolution of the meridional flow measured by magnetic pattern tracking of MDI/HMI data over the last two cycles is shown in the left panel (adapted from \citet{2022Hathaway_etal}). 
    }
    \label{fig:lu_mc2}
\end{figure}

\begin{figure}[th!]
    \centering
        \includegraphics[trim=0.0in 0.in -0.50in .20in, clip, width=0.95\linewidth]{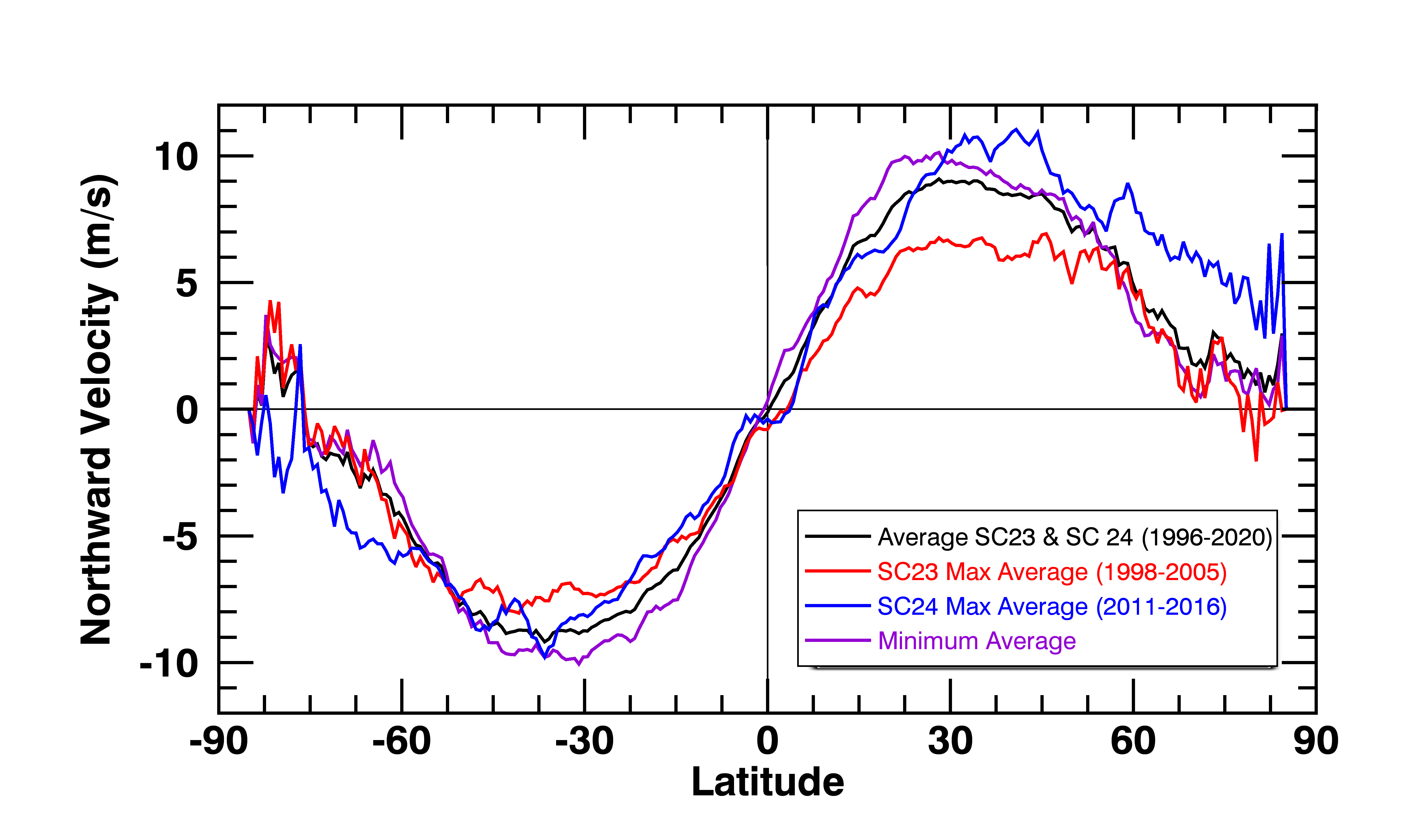}
        \includegraphics[trim=3.6in 0.in 0.0in 0.0in, clip, width=0.75\linewidth]{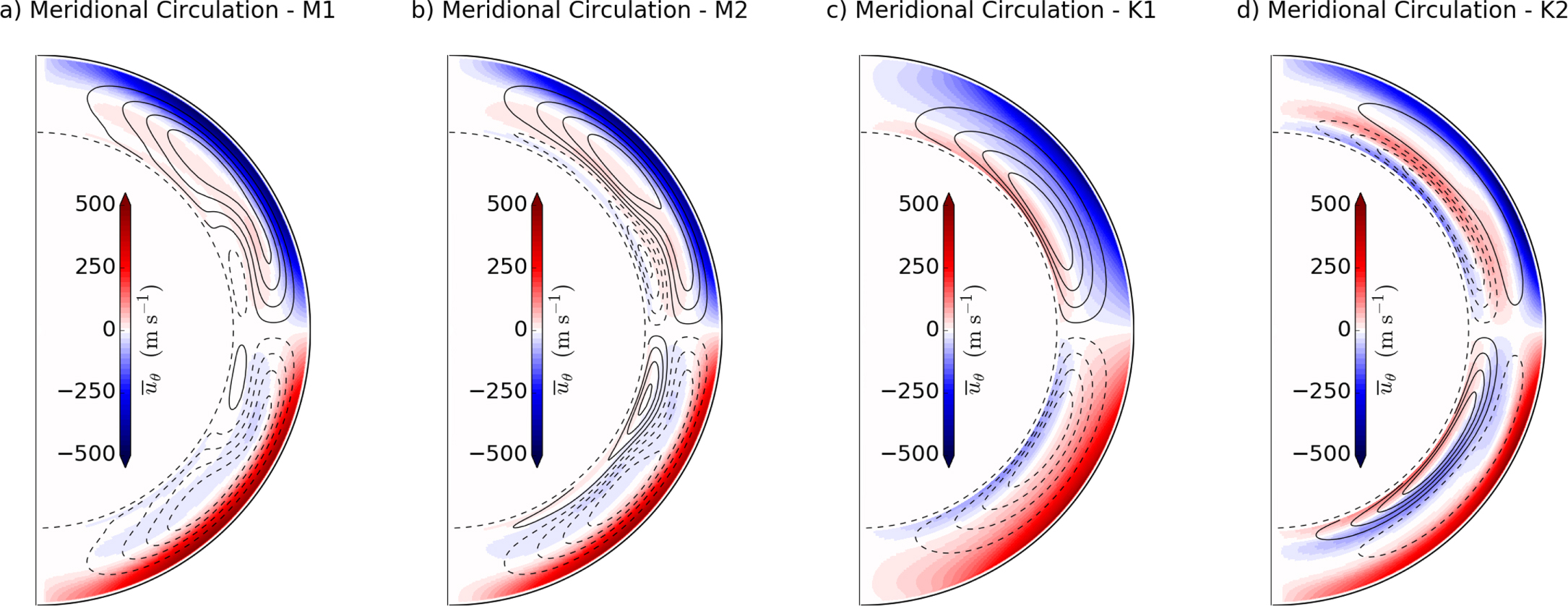}

    \caption{The average meridional flow profiles for different time periods measured by magnetic pattern tracking of MDI/HMI data are shown in the top panel. The red (blue) represents the flow during cycle 23 (24) maximum. The purple line represents cycle minimum during Solar Cycle 23 and 24. The black line represents an average over both cycles.  Two possible, idealized meridional circulation patterns are shown in the bottom panel, adapted from \citet{stejko:2021} (reproduced by permission): the classical single-cell with a deep return flow (labeled K1) and a double-cell circulation profile with a stronger return flow (labeled K2).}
    \label{fig:lu_mc3}
\end{figure}

High resolution continuous magnetic data from space-based observatories (i.e., SOHO/MDI and SDO/HMI) have ushered in a new era, paving the way for meridional flow measurements with unprecedented spatial and temporal resolution, revealing that the amplitude and structure vary with the solar cycle \citep{2004Gizon, 2008GonzalezHernandez_etal, 2010HathawayRightmire}. The meridional flow measured by magnetic pattern tracking \citep{2022Hathaway_etal} for the last two solar cycles (see Figure~\ref{fig:lu_mc2}) shows that the meridional flow is the strongest at solar cycle minimum and weakens during solar minimum. This weakening of the meridional flow was more pronounced during the stronger Solar Cycle 23 than it was for the weak Solar Cycle 24. The relative magnitude of this cycle dependent change in the flow speed is illustrated in the left panel of Figure~\ref{fig:lu_mc3}. This modulation of the meridional flow by the presence of ARs may serve as a nonlinear feedback mechanism for regulating the solar cycle, as described in the next section.

Another aspect of the meridional flow is the quite contentious existence of the  high-latitude equatorward flows, sometimes referred to as polar counter-cells. If present, these flows would have implications for the build up of the polar fields and thus the strength of the solar cycle \citep{Jiang:2009, Upton:2014b}. The possibility of these flows was suggested by \citet{Ulrich_2010} as well as \citet{2010HathawayRightmire}, but later dismissed \citep{2012RightmireUpton_etal} as an instrumental artifact because the counter-cells were not originally present in high-resolution HMI data. However, more recent analysis \citep{2022Hathaway_etal} now suggests these flows may have returned and are now observed in the HMI measurements (see Figure~\ref{fig:lu_mc2}). As of yet, their appearance does not seem to have a solar-cycle dependence but rather to occur somewhat sporadically. Resolving these structures unambiguously remains a challenge for several reasons. First and foremost, these flows only appear to be $\sim 1-2$ m s$^{-1}$, an order of magnitude weaker than the already difficult to measure standard meridional flow. Secondly, they appear at latitudes above 60$^\circ$, where the radial component of the magnetic field is not well resolved and signal to noise is small. While advancement in the measurement techniques may eventually shed some light on this ambiguous aspect of the meridional flow, a mission to directly observe the poles with a Doppler-magnetograph may ultimately be needed to fully resolve these controversial flows.   

In order to satisfy mass conservation, the meridional flow must have an equatorward return flow at some depth, and thus it is also referred to as the meridional circulation. In addition to generating the polar fields at the surface to initialize the solar cycle, the meridional circulation in the interior is believed to play an important role in setting the period of the cycle \citep{dikpati:1999b}. Long thought to be a single circulating cell in each hemisphere, modern observations are challenging that notion \citep{2012Hathaway, 2013Zhao_etal} with indications that a double cell may exist at times in each hemisphere. However, \citet{gizon:2020} finds evidence of a single meridional circulation cell using recent observations, so a discrepancy exists. Understanding the implications of different possible configurations (e.g., see the lower panel of Figure~\ref{fig:lu_mc3}) of the meridional return flow in the solar interior has become an integral focus of dynamo modelers \citep{2017BekkiYokoyama, stejko:2021}. For a more in depth discussion on this, refer to \citet{hazra:2023, hotta:2023}.

\subsection{Active Region Inflows}

Inflows towards AR belts are observed by local helioseismic techniques (e.g., \citep{gizon:2001, zhao:2004, haber:2004, gonzalez:2008} and these flows are observed from approximately 10$^{\circ}$ from the AR with amplitudes up to 50 m s$^{-1}$ of horizontal velocities.  The AR inflows modulate the N-S meridional flow, which is on the same order of magnitude.  

The explanation for the inflows is a geostrophic flow caused by increased radiative loss in the AR belt \citep{gizon:2008}. \citet{gottschling:2022} studied the evolution of the AR inflows and reports that converging flows are present one day prior to emergence and that these pre-emergence flows do not depend on latitude or flux. A prograde flow of about 40\,m s$^{-1}$ is found at the leading polarity during emergence \citep{birch:2019, gottschling:2022} with the increase in amplitudes of the inflows occurring between 1$-$4 days after emergence. 

One important consequence of AR inflows is that they slow the flux diffusion, advection, and cancellation \citep{derosa:2006}. Surface flux transport modeling highlights how AR inflows may modulate the amplitude of the global magnetic field in several ways. First, AR inflows can limit the latitudinal separation of the AR polarities,
thus weakening of the contribution of a given bipolar
region to the axial dipole field \citep{jiang:2010}, or second, by increasing the cross-equatorial transport of magnetic flux in weaker cycles, when sunspots emerge at lower latitudes, which ultimately strengthens the axial dipole field \citep{cameron:2012}.  For further discussion of the implementation of AR inflows into surface flux transport models, and the results thereof, see \citet{yeates:2023} discussion on fluctuating large-scale flows. 

\section{Short-Term Solar Cycle Variability}\label{Short-term}
There are two significant variations seen in solar-cycle data (i.e., sunspot number and area, 10.7 cm radio emission, mean solar magnetic field, coronal green line, H-alpha flare number, solar neutrino flux, p-mode frequencies) on a time period shorter than the sunspot cycle: the quasi-biennial oscillation (QBO) (1$-$4 years) and Rieger-type variations (50$-$200 days). The QBO was observed as a roughly 2-year period \citep{benevolenskaya:1995} in polar field components and also manifests as the double peak in sunspot numbers observed most easily near the maximum of the sunspot cycle.  This double-peak is also known as the Gnevyshev gap \citep{gnevyshev:1967}. A thorough look at QBOs can be found in \citet{bazilevskaya:2014} in which the following characteristics are listed: the QBO timescales change within the range 1\,--\,4 years with no dominant frequency; they develop in each solar hemisphere independently, but are synchronous within one hemisphere with signatures in the atmosphere and beneath the photosphere based on helioseismology \citep{bazilevskaya:2014}; they are observed in the photospheric magnetic field in phase with other solar activity indices; the QBOs are transferred into the interplanetary medium by the Sun’s open magnetic flux. The Rieger type variations were first observed in gamma-ray flare activity in the 1980s with a 154 d periodicity \citep{rieger:1984}. They were subsequently shown to have many shorter periodicities and be present in sunspot number and area and photospheric magnetic field indices \citep{bai:2003}.

Other stars show secondary, shorter cycles with smaller amplitudes than their primary cycle \citep{bohm:2007}, and one explanation is that the dynamo is fed by the deep-seated and near-surface shear layer. Another explanation is that the QBOs are caused by the interaction between the dipole and quadrupole terms of the solar dynamo\citep{wang:2003}. The physical mechanisms responsible for the Rieger type variability may be as simple as AR evolution \citep{vecchio:2012} or a harmonic of the QBO \citep{krivova:2002}. One compelling mechanism that can produce a range of short-term variability is found in MHD shallow-water modeling of an instability involving the differential rotation and toroidal field bands in the solar tachocline \citep{dikpati:2005}. This instability generates quasi-periodic tachocline nonlinear oscillations (TNOs) with periodicities of 2-20 months that can be correlated with the formation of persistent active longitudes \citep{detoma:2000} seen in photospheric magnetic field data. \citep{dikpati:2018, dikpati:2021}. The mechanism involves the production of upward bulges at selected longitudes in the overshoot tachocline that contain significant toroidal fields. For a review of long-term modulation of the solar cycle, see \citet{biswas:2023}.

\section{Helioseismic Mode Parameter Changes}\label{Mode}

The frequency \citep[e.g.][]{1985Natur.318..449W,1990Natur.345..779L,1990Natur.345..322E}, amplitude \citep[e.g.][]{1993MNRAS.265..888E}, and lifetime\citep{2000MNRAS.313...32C, 2002ApJ...572..663K} of the acoustic modes used in helioseismology all vary with the solar cycle, and they are spatially and temporally correlated with magnetic activity on a wide range of scales, with changes in low-degree modes following global activity measures such as the sunspot number and 10.7\,cm radio flux (RF), while in local helioseismology we can see changes down to the scale of ARs \citep{2000SoPh..192..363H, 2001ApJ...563..410R, 2004ApJ...608..562H}. Most of these changes are believed \citep[e.g.][]{1990Natur.345..779L}to arise quite close to the surface, where the cavity in which the modes propagate is modified directly or indirectly by the presence of activity and the excitation and damping of the modes influenced by magnetic fields. The exact interpretation of these changes is difficult, but, for example 

\cite{2004ApJ...617L.155B}, \cite{2006ApJ...640L..95V}, and more recently \cite{2020ApJ...903L..29W} found evidence of solar-cycle changes in the signature that the helium ionization zone at 0.98\,${\mathrm{R}_{\odot}}$ makes in helioseismic frequencies.
These near-surface effects dominate the changes and make it difficult to use helioseismology to infer changes in the internal solar structure or magnetic fields.  

While short-lived, high-degree modes can be used to study local, near-surface effects on timescales as short as a day, global helioseismology requires integration times of at least a few solar rotations to obtain the necessary precision to resolve the interior structure and dynamics, and this precludes the possibility, for example, of using helioseismology to follow the rise of an individual flux tube through the convection zone. On the timescale of a solar cycle, some marginal effects have been reported. For example, \cite{2008ApJ...686.1349B} and more recently \cite{2021ApJ...917...45B} found small changes in the sound speed at the base of the convection zone. The latter work reports a change of about $2\times 10^{-5}$ in the squared sound-speed at the base of the convection zone between solar maximum and minimum, anticorrelated with the activity level; this is just below the $3\times10^{-5}$ upper limit found by \cite{2002ApJ...580..574E}. Small changes in the sound-speed near the base of the convection zone were also seen by \citet{2005ApJ...624..420C} using a different technique. Because these effects are so difficult to measure, they have not been widely studied. Any effects of solar-cycle changes in the magnetic fields near the base of the convection zone, which could be valuable to help in understanding the solar dynamo, remain close to or below the limits of detection.

\section{Discussion}

The traditional index of solar activity is the (group) sunspot number, which however, is robustly defined before the middle of the 19th century and particularly poor in the first half of the 18th century. The research community is working hard to reconcile the sunspot dataset.

While the solar-cycle phenomena in the Sun's surface and outer atmosphere can be studied in great detail using a variety of observing techniques, helioseismology reveals motions -- and to a limited extent structural changes related to the solar cycle -- far below the photosphere. The torsional oscillation and the meridional circulation penetrate throughout the convection zone and play a crucial role in the solar dynamo. Therefore, it is imperative that we achieve an unambiguous inference of the structure of the meridional circulation at depth and its evolution over the solar cycle. If, for example, the observations completely ruled out a single-cell meridional circulation flow in each hemisphere, it would  rule out dynamo models that rely on such a configuration.

We are able to infer flows at depths, yet the great majority of solar magnetic fields remain unobserved in the solar interior. The bipolar magnetic regions observed in the photosphere represent only the ``tip of the iceberg". Until we can reliably infer magnetic field strengths and dynamics in the interior, we must rely on observations at the surface coupled with simulations to infer dynamics, amplitude and structure of the magnetic fields at depth. 

To be considered successful, solar dynamo theories and simulations must be able to reproduce, to some degree, key observations. This includes the modulation of cycle amplitude as measured in the sunspot number, the observed large-scale flows, the adherence to Hale's law, the trends and inherent scatter in the tilt angles, the equator-ward migration of the active latitudes that produces the butterfly diagram, the evolution of the polar fields. 


\section*{Declarations}
\backmatter
\bmhead{Funding}
I.U. acknowledges partial support from the Academy of Finland (project 321882 ESPERA). L.U. was supported by NASA Heliophysics Living With a Star grants 80HQTR18T0023, 80NSSC20K0220, and 80NSSC23K0048 and NASA grant 80NSSC22M0162 to the COFFIES DRIVE Center managed by Stanford University. A.N. acknowledges NASA DRIVE Center COFFIES grant 80NSSC20K0602.
R.H. acknowledges the support of the UK Science and Technology Facilities Council (STFC) through
grant ST/V000500/1.
\bmhead{Author Contributions}
I.U. contributed $\mathsection$2 and Figure 1. A.N. contributed $\mathsection$1, 3, 6.4, 7 and Figures 2 and 3. L.U. contributed $\mathsection$4, 5 and Figures 4, 5, 6, 10 and 11. R.H. contributed $\mathsection$6.1-3, 8 and Figures 7, 8, and 9. All authors reviewed the full manuscript and contributed to $\mathsection$9. 

\bmhead{Conflicts of Interest}
The authors have no relevant financial or non-financial conflicts of interest to disclose.

\bmhead{Acknowledgments} 
All authors thank the International Space Science Institute for supporting the workshop where this review originated. 

\bibliography{SCObs}{}

\begin{thebibliography}{157}
\providecommand{\natexlab}[1]{#1}
\providecommand{\url}[1]{{#1}}
\providecommand{\urlprefix}{URL }
\providecommand{\doi}[1]{\url{https://doi.org/#1}}
\providecommand{\eprint}[2][]{\url{#2}}
 \bibcommenthead

\bibitem[{{Antia} and {Basu}(2001)}]{2001ApJ...559L..67A}
{Antia} HM, {Basu} S (2001) {Temporal Variations of the Solar Rotation Rate at
  High Latitudes}. \apjl 559(1):L67--L70. \doi{10.1086/323701},
  {\href{https://arxiv.org/abs/astro-ph/0108226}{{https://arxiv.org/abs/arXiv:astro-ph/0108226}}}
  {[astro-ph]}

\bibitem[{{Asvestari} et~al(2017){Asvestari}, {Usoskin}, {Kovaltsov}, {Owens},
  {Krivova}, {Rubinetti}, and {Taricco}}]{asvestari_MNRAS_17}
{Asvestari} E, {Usoskin} IG, {Kovaltsov} GA, et~al (2017) {Assessment of
  different sunspot number series using the cosmogenic isotope $^{44}$Ti in
  meteorites}. Monthly Notes Royal Astro Soc 467:1608--1613.
  \doi{10.1093/mnras/stx190}

\bibitem[{{Babcock}(1961)}]{babcock:1961}
{Babcock} HW (1961) {The Topology of the Sun's Magnetic Field and the 22-YEAR
  Cycle.} \apj 133:572. \doi{10.1086/147060}

\bibitem[{{Bai}(2003)}]{bai:2003}
{Bai} T (2003) {Periodicities in Solar Flare Occurrence: Analysis of Cycles
  19-23}. \apj 591(1):406--415. \doi{10.1086/375295}

\bibitem[{{Baldner} and {Basu}(2008)}]{2008ApJ...686.1349B}
{Baldner} CS, {Basu} S (2008) {Solar Cycle Related Changes at the Base of the
  Convection Zone}. \apj 686(2):1349--1361. \doi{10.1086/591514},
  {\href{https://arxiv.org/abs/0807.0442}{{https://arxiv.org/abs/arXiv:0807.0442}}}
  {[astro-ph]}

\bibitem[{{Basu}(2021)}]{2021ApJ...917...45B}
{Basu} S (2021) {Evidence of Solar-cycle-related Structural Changes in the
  Solar Convection Zone}. \apj 917(1):45. \doi{10.3847/1538-4357/ac0c11},
  {\href{https://arxiv.org/abs/2106.08383}{{https://arxiv.org/abs/arXiv:2106.08383}}}
  {[astro-ph.SR]}

\bibitem[{{Basu} and {Mandel}(2004)}]{2004ApJ...617L.155B}
{Basu} S, {Mandel} A (2004) {Does Solar Structure Vary with Solar Magnetic
  Activity?} \apjl 617(2):L155--L158. \doi{10.1086/427435},
  {\href{https://arxiv.org/abs/astro-ph/0411427}{{https://arxiv.org/abs/arXiv:astro-ph/0411427}}}
  {[astro-ph]}

\bibitem[{{Bazilevskaya} et~al(2014){Bazilevskaya}, {Broomhall}, {Elsworth},
  and {Nakariakov}}]{bazilevskaya:2014}
{Bazilevskaya} G, {Broomhall} AM, {Elsworth} Y, et~al (2014) {A Combined
  Analysis of the Observational Aspects of the Quasi-biennial Oscillation in
  Solar Magnetic Activity}. \ssr 186(1-4):359--386.
  \doi{10.1007/s11214-014-0068-0}

\bibitem[{{Beck}(2000)}]{2000SoPh..191...47B}
{Beck} JG (2000) {A comparison of differential rotation measurements - (Invited
  Review)}. \solphys 191(1):47--70. \doi{10.1023/A:1005226402796}

\bibitem[{{Bekki} and {Yokoyama}(2017)}]{2017BekkiYokoyama}
{Bekki} Y, {Yokoyama} T (2017) {Double-cell-type Solar Meridional Circulation
  Based on a Mean-field Hydrodynamic Model}. \apj 835(1):9.
  \doi{10.3847/1538-4357/835/1/9},
  {\href{https://arxiv.org/abs/1612.00174}{{https://arxiv.org/abs/arXiv:1612.00174}}}
  {[astro-ph.SR]}

\bibitem[{{Benevolenskaya}(1995)}]{benevolenskaya:1995}
{Benevolenskaya} EE (1995) {Double Magnetic Cycle of Solar Activity}. \solphys
  161(1):1--8. \doi{10.1007/BF00732080}

\bibitem[{{Bhowmik} et~al(2023){Bhowmik}, {Jiang}, {Upton}, {Lemerle}, and
  {Nandy}}]{bhowmik:2023}
{Bhowmik} P, {Jiang} J, {Upton} L, et~al (2023) {Physical Models for Solar
  Cycle Predictions}. \ssr 219(5):40. \doi{10.1007/s11214-023-00983-x},
  {\href{https://arxiv.org/abs/2303.12648}{{https://arxiv.org/abs/arXiv:2303.12648}}}
  {[astro-ph.SR]}

\bibitem[{{Birch} et~al(2019){Birch}, {Schunker}, {Braun}, and
  {Gizon}}]{birch:2019}
{Birch} AC, {Schunker} H, {Braun} DC, et~al (2019) {Average surface flows
  before the formation of solar active regions and their relationship to the
  supergranulation pattern}. \aap 628:A37. \doi{10.1051/0004-6361/201935591}

\bibitem[{{Biswas} et~al(2023){Biswas}, {Karak}, {Usoskin}, and
  {Weisshaar}}]{biswas:2023}
{Biswas} A, {Karak} BB, {Usoskin} I, et~al (2023) {Long-Term Modulation of
  Solar Cycles}. \ssr 219(3):19. \doi{10.1007/s11214-023-00968-w},
  {\href{https://arxiv.org/abs/2302.14845}{{https://arxiv.org/abs/arXiv:2302.14845}}}
  {[astro-ph.SR]}

\bibitem[{{B{\"o}hm-Vitense}(2007)}]{bohm:2007}
{B{\"o}hm-Vitense} E (2007) {Chromospheric Activity in G and K Main-Sequence
  Stars, and What It Tells Us about Stellar Dynamos}. \apj 657(1):486--493.
  \doi{10.1086/510482}

\bibitem[{{Cameron} and {Sch{\"u}ssler}(2023)}]{cameron:2023}
{Cameron} R, {Sch{\"u}ssler} M (2023) {Observationally guided models for the
  solar dynamo and the role of the surface field}. arXiv e-prints
  arXiv:2305.02253. \doi{10.48550/arXiv.2305.02253},
  {\href{https://arxiv.org/abs/2305.02253}{{https://arxiv.org/abs/arXiv:2305.02253}}}
  {[astro-ph.SR]}

\bibitem[{{Cameron} and {Sch{\"u}ssler}(2012)}]{cameron:2012}
{Cameron} RH, {Sch{\"u}ssler} M (2012) {Are the strengths of solar cycles
  determined by converging flows towards the activity belts?} \aap 548:A57.
  \doi{10.1051/0004-6361/201219914},
  {\href{https://arxiv.org/abs/1210.7644}{{https://arxiv.org/abs/arXiv:1210.7644}}}
  {[astro-ph.SR]}

\bibitem[{{Cameron} et~al(2018){Cameron}, {Duvall}, {Sch{\"u}ssler}, and
  {Schunker}}]{cameron:2018}
{Cameron} RH, {Duvall} TL, {Sch{\"u}ssler} M, et~al (2018) {Observing and
  modeling the poloidal and toroidal fields of the solar dynamo}. \aap 609:A56.
  \doi{10.1051/0004-6361/201731481},
  {\href{https://arxiv.org/abs/1710.07126}{{https://arxiv.org/abs/arXiv:1710.07126}}}
  {[astro-ph.SR]}

\bibitem[{{Carrasco} et~al(2021){Carrasco}, {Hayakawa}, {Kuroyanagi},
  {Gallego}, and {Vaquero}}]{carrasco21}
{Carrasco} VMS, {Hayakawa} H, {Kuroyanagi} C, et~al (2021) {Strong evidence of
  low levels of solar activity during the Maunder Minimum}. Mon Not R Astron
  Soc 504(4):5199--5204. \doi{10.1093/mnras/stab1155}

\bibitem[{{Chaplin} et~al(2000){Chaplin}, {Elsworth}, {Isaak}, {Miller}, and
  {New}}]{2000MNRAS.313...32C}
{Chaplin} WJ, {Elsworth} Y, {Isaak} GR, et~al (2000) {Variations in the
  excitation and damping of low-l solar p modes over the solar activity
  cycle$^{*}$}. \mnras 313(1):32--42. \doi{10.1046/j.1365-8711.2000.03176.x}

\bibitem[{{Charbonneau} and {Sokoloff}(2023)}]{charbonneau:2023}
{Charbonneau} P, {Sokoloff} D (2023) {Evolution of Solar and Stellar Dynamo
  Theory}. \ssr 219(5):35. \doi{10.1007/s11214-023-00980-0},
  {\href{https://arxiv.org/abs/2305.16553}{{https://arxiv.org/abs/arXiv:2305.16553}}}
  {[astro-ph.SR]}

\bibitem[{{Charbonneau} et~al(2005){Charbonneau}, {St-Jean}, and
  {Zacharias}}]{char:2005}
{Charbonneau} P, {St-Jean} C, {Zacharias} P (2005) {Fluctuations in
  Babcock-Leighton Dynamos. I. Period Doubling and Transition to Chaos}. \apj
  619(1):613--622. \doi{10.1086/426385}

\bibitem[{{Chatzistergos} et~al(2017){Chatzistergos}, {Usoskin}, {Kovaltsov},
  {Krivova}, and {Solanki}}]{chatzistergos17}
{Chatzistergos} T, {Usoskin} IG, {Kovaltsov} GA, et~al (2017) {New
  reconstruction of the sunspot group numbers since 1739 using direct
  calibration and ``backbone'' methods}. Astron Astrophys 602:A69.
  \doi{10.1051/0004-6361/201630045},
  {\href{https://arxiv.org/abs/1702.06183}{{https://arxiv.org/abs/arXiv:1702.06183}}}
  {[astro-ph.SR]}

\bibitem[{{Chou} and {Serebryanskiy}(2005)}]{2005ApJ...624..420C}
{Chou} DY, {Serebryanskiy} A (2005) {In Search of the Solar Cycle Variations of
  p-Mode Frequencies Generated by Perturbations in the Solar Interior}. \apj
  624:420--427. \doi{10.1086/428925},
  {\href{https://arxiv.org/abs/astro-ph/0405175}{{https://arxiv.org/abs/astro-ph/0405175}}}

\bibitem[{{Clette} and {Lef{\`e}vre}(2016)}]{clette16}
{Clette} F, {Lef{\`e}vre} L (2016) {The New Sunspot Number: Assembling All
  Corrections}. Solar Phys 291:2629--2651. \doi{10.1007/s11207-016-1014-y}

\bibitem[{{Clette} et~al(2007){Clette}, {Berghmans}, {Vanlommel}, {Van der
  Linden}, {Koeckelenbergh}, and {Wauters}}]{clette07}
{Clette} F, {Berghmans} D, {Vanlommel} P, et~al (2007) {From the Wolf number to
  the International Sunspot Index: 25 years of SIDC}. Adv Space Res
  40:919--928. \doi{10.1016/j.asr.2006.12.045}

\bibitem[{Clette et~al(2014)Clette, Svalgaard, Vaquero, and Cliver}]{clette14}
Clette F, Svalgaard L, Vaquero J, et~al (2014) Revisiting the sunspot number: A
  400-year perspective on the solar cycle. Space Sci Rev 186:35.
  \doi{10.1007/s11214-014-0074-2}

\bibitem[{{Corbard} et~al(1999){Corbard}, {Blanc-F{\'e}raud}, {Berthomieu}, and
  {Provost}}]{1999A&A...344..696C}
{Corbard} T, {Blanc-F{\'e}raud} L, {Berthomieu} G, et~al (1999) {Non linear
  regularization for helioseismic inversions. Application for the study of the
  solar tachocline}. \aap 344:696--708.
  {\href{https://arxiv.org/abs/astro-ph/9901112}{{https://arxiv.org/abs/arXiv:astro-ph/9901112}}}
  {[astro-ph]}

\bibitem[{{Couvidat} et~al(2003){Couvidat}, {Garc{\'\i}a}, {Turck-Chi{\`e}ze},
  {Corbard}, {Henney}, and {Jim{\'e}nez-Reyes}}]{2003ApJ...597L..77C}
{Couvidat} S, {Garc{\'\i}a} RA, {Turck-Chi{\`e}ze} S, et~al (2003) {The
  Rotation of the Deep Solar Layers}. \apjl 597(1):L77--L79.
  \doi{10.1086/379698},
  {\href{https://arxiv.org/abs/astro-ph/0309806}{{https://arxiv.org/abs/arXiv:astro-ph/0309806}}}
  {[astro-ph]}

\bibitem[{{Dasi-Espuig} et~al(2010){Dasi-Espuig}, {Solanki}, {Krivova},
  {Cameron}, and {Pe{\~n}uela}}]{dasi-espuig:2010}
{Dasi-Espuig} M, {Solanki} SK, {Krivova} NA, et~al (2010) {Sunspot group tilt
  angles and the strength of the solar cycle}. \aap 518:A7.
  \doi{10.1051/0004-6361/201014301},
  {\href{https://arxiv.org/abs/1005.1774}{{https://arxiv.org/abs/arXiv:1005.1774}}}
  {[astro-ph.SR]}

\bibitem[{{De Rosa} and {Schrijver}(2006)}]{derosa:2006}
{De Rosa} ML, {Schrijver} CJ (2006) {Consequences of large-scale flows around
  active regions on the dispersal of magnetic field across the solar surface}.
  In: {Fletcher} K, {Thompson} M (eds) Proceedings of SOHO 18/GONG 2006/HELAS
  I, Beyond the spherical Sun, p~12

\bibitem[{{de Toma} et~al(2000){de Toma}, {White}, and {Harvey}}]{detoma:2000}
{de Toma} G, {White} OR, {Harvey} KL (2000) {A Picture of Solar Minimum and the
  Onset of Solar Cycle 23. I. Global Magnetic Field Evolution}. \apj
  529(2):1101--1114. \doi{10.1086/308299}

\bibitem[{{Dikpati} and {Charbonneau}(1999{\natexlab{a}})}]{dikpati:1999}
{Dikpati} M, {Charbonneau} P (1999{\natexlab{a}}) {A Babcock-Leighton Flux
  Transport Dynamo with Solar-like Differential Rotation}. \apj
  518(1):508--520. \doi{10.1086/307269}

\bibitem[{{Dikpati} and {Charbonneau}(1999{\natexlab{b}})}]{dikpati:1999b}
{Dikpati} M, {Charbonneau} P (1999{\natexlab{b}}) {Intermittency in Solar Cycle
  Caused by Stochastic Fluctuation in Meridional Circulation}. In: American
  Astronomical Society Meeting Abstracts \#194, p 92.04

\bibitem[{{Dikpati} and {Gilman}(2005)}]{dikpati:2005}
{Dikpati} M, {Gilman} PA (2005) {A Shallow-Water Theory for the Sun's Active
  Longitudes}. \apjl 635(2):L193--L196. \doi{10.1086/499626}

\bibitem[{{Dikpati} et~al(2018){Dikpati}, {McIntosh}, {Bothun}, {Cally},
  {Ghosh}, {Gilman}, and {Umurhan}}]{dikpati:2018}
{Dikpati} M, {McIntosh} SW, {Bothun} G, et~al (2018) {Role of Interaction
  between Magnetic Rossby Waves and Tachocline Differential Rotation in
  Producing Solar Seasons}. \apj 853(2):144. \doi{10.3847/1538-4357/aaa70d}

\bibitem[{{Dikpati} et~al(2021){Dikpati}, {McIntosh}, {Chatterjee}, {Norton},
  {Ambroz}, {Gilman}, {Jain}, and {Munoz-Jaramillo}}]{dikpati:2021}
{Dikpati} M, {McIntosh} SW, {Chatterjee} S, et~al (2021) {Deciphering the Deep
  Origin of Active Regions via Analysis of Magnetograms}. \apj 910(2):91.
  \doi{10.3847/1538-4357/abe043}

\bibitem[{{D'Silva} and {Choudhuri}(1993)}]{d'silva:1993}
{D'Silva} S, {Choudhuri} AR (1993) {A theoretical model for tilts of bipolar
  magnetic regions}. \aap 272:621

\bibitem[{Eddy(1976)}]{eddy76}
Eddy J (1976) The maunder minimum. Science 192:1189--1202.
  \doi{10.1126/science.192.4245.1189}

\bibitem[{{Eff-Darwich} and {Korzennik}(1998)}]{1998ESASP.418..685E}
{Eff-Darwich} A, {Korzennik} SG (1998) {The Rotation of the Solar Core:
  Compatibility of the Different Data Sets Available}. In: {Korzennik} S (ed)
  Structure and Dynamics of the Interior of the Sun and Sun-like Stars, p 685

\bibitem[{{Eff-Darwich} et~al(2002){Eff-Darwich}, {Korzennik},
  {Jim{\'e}nez-Reyes}, and {P{\'e}rez Hern{\'a}ndez}}]{2002ApJ...580..574E}
{Eff-Darwich} A, {Korzennik} SG, {Jim{\'e}nez-Reyes} SJ, et~al (2002) {An Upper
  Limit on the Temporal Variations of the Solar Interior Stratification}. \apj
  580(1):574--578. \doi{10.1086/343063},
  {\href{https://arxiv.org/abs/astro-ph/0207402}{{https://arxiv.org/abs/arXiv:astro-ph/0207402}}}
  {[astro-ph]}

\bibitem[{{Elsworth} et~al(1990){Elsworth}, {Howe}, {Isaak}, {McLeod}, and
  {New}}]{1990Natur.345..322E}
{Elsworth} Y, {Howe} R, {Isaak} GR, et~al (1990) {Variation of low-order
  acoustic solar oscillations over the solar cycle}. \nat 345:322--324.
  \doi{10.1038/345322a0}

\bibitem[{{Elsworth} et~al(1993){Elsworth}, {Howe}, {Isaak}, {McLeod},
  {Miller}, {Speake}, {Wheeler}, and {New}}]{1993MNRAS.265..888E}
{Elsworth} Y, {Howe} R, {Isaak} GR, et~al (1993) {The variation in the strength
  of low-l solar p-modes - 1981-92}. \mnras 265:888--898

\bibitem[{{Fan}(2009)}]{fan:2009}
{Fan} Y (2009) {Magnetic Fields in the Solar Convection Zone}. Living Reviews
  in Solar Physics 6(1):4. \doi{10.12942/lrsp-2009-4}

\bibitem[{{Fisher} et~al(1995){Fisher}, {Fan}, and {Howard}}]{fisher:1995}
{Fisher} GH, {Fan} Y, {Howard} RF (1995) {Comparisons between Theory and
  Observation of Active Region Tilts}. \apj 438:463. \doi{10.1086/175090}

\bibitem[{{Friedli}(2020)}]{friedli20}
{Friedli} TK (2020) {Recalculation of the Wolf Series from 1877 to 1893}. Solar
  Phys 295(6):72. \doi{10.1007/s11207-020-01637-9}

\bibitem[{{Gilman} and {Howard}(1986)}]{gilman:1986}
{Gilman} PA, {Howard} R (1986) {Rotation and Expansion within Sunspot Groups}.
  \apj 303:480. \doi{10.1086/164093}

\bibitem[{{Gilman} and {Howe}(2003)}]{2003soho...12..283G}
{Gilman} PA, {Howe} R (2003) {Meridional motion and the slope of isorotation
  contours}. In: {Sawaya-Lacoste} H (ed) ESA SP-517: GONG+ 2002. Local and
  Global Helioseismology: the Present and Future, pp 283--285

\bibitem[{{Gizon}(2004)}]{2004Gizon}
{Gizon} L (2004) {Helioseismology of Time-Varying Flows Through The Solar
  Cycle}. \solphys 224(1-2):217--228. \doi{10.1007/s11207-005-4983-9}

\bibitem[{{Gizon} and {Rempel}(2008)}]{gizon:2008}
{Gizon} L, {Rempel} M (2008) {Observation and Modeling of the Solar-Cycle
  Variation of the Meridional Flow}. \solphys 251(1-2):241--250.
  \doi{10.1007/s11207-008-9162-3},
  {\href{https://arxiv.org/abs/0803.0950}{{https://arxiv.org/abs/arXiv:0803.0950}}}
  {[astro-ph]}

\bibitem[{{Gizon} et~al(2001){Gizon}, {Duvall}, and {Larsen}}]{gizon:2001}
{Gizon} L, {Duvall} JT.~L., {Larsen} RM (2001) {Probing Surface Flows and
  Magnetic Activity with Time-Distance Helioseismology}. In: {Brekke} P,
  {Fleck} B, {Gurman} JB (eds) Recent Insights into the Physics of the Sun and
  Heliosphere: Highlights from SOHO and Other Space Missions, p 189

\bibitem[{{Gizon} et~al(2020){Gizon}, {Cameron}, {Pourabdian}, {Liang},
  {Fournier}, {Birch}, and {Hanson}}]{gizon:2020}
{Gizon} L, {Cameron} RH, {Pourabdian} M, et~al (2020) {Meridional flow in the
  Sun{\textquoteright}s convection zone is a single cell in each hemisphere}.
  Science 368(6498):1469--1472. \doi{10.1126/science.aaz7119}

\bibitem[{{Gnevyshev}(1967)}]{gnevyshev:1967}
{Gnevyshev} MN (1967) {On the 11-Years Cycle of Solar Activity}. \solphys
  1(1):107--120. \doi{10.1007/BF00150306}

\bibitem[{{Gonz{\'a}lez Hern{\'a}ndez} et~al(2008{\natexlab{a}}){Gonz{\'a}lez
  Hern{\'a}ndez}, {Kholikov}, {Hill}, {Howe}, and
  {Komm}}]{2008GonzalezHernandez_etal}
{Gonz{\'a}lez Hern{\'a}ndez} I, {Kholikov} S, {Hill} F, et~al
  (2008{\natexlab{a}}) {Subsurface Meridional Circulation in the Active Belts}.
  \solphys 252(2):235--245. \doi{10.1007/s11207-008-9264-y},
  {\href{https://arxiv.org/abs/0808.3606}{{https://arxiv.org/abs/arXiv:0808.3606}}}
  {[astro-ph]}

\bibitem[{{Gonz{\'a}lez Hern{\'a}ndez} et~al(2008{\natexlab{b}}){Gonz{\'a}lez
  Hern{\'a}ndez}, {Kholikov}, {Hill}, {Howe}, and {Komm}}]{gonzalez:2008}
{Gonz{\'a}lez Hern{\'a}ndez} I, {Kholikov} S, {Hill} F, et~al
  (2008{\natexlab{b}}) {Subsurface Meridional Circulation in the Active Belts}.
  \solphys 252(2):235--245. \doi{10.1007/s11207-008-9264-y},
  {\href{https://arxiv.org/abs/0808.3606}{{https://arxiv.org/abs/arXiv:0808.3606}}}
  {[astro-ph]}

\bibitem[{{Gottschling} et~al(2022){Gottschling}, {Schunker}, {Birch},
  {Cameron}, and {Gizon}}]{gottschling:2022}
{Gottschling} N, {Schunker} H, {Birch} AC, et~al (2022) {Testing solar surface
  flux transport models in the first days after active region emergence}. \aap
  660:A6. \doi{10.1051/0004-6361/202142071},
  {\href{https://arxiv.org/abs/2111.01896}{{https://arxiv.org/abs/arXiv:2111.01896}}}
  {[astro-ph.SR]}

\bibitem[{{Haber} et~al(2004){Haber}, {Hindman}, {Toomre}, and
  {Thompson}}]{haber:2004}
{Haber} DA, {Hindman} BW, {Toomre} J, et~al (2004) {Organized Subsurface Flows
  near Active Regions}. \solphys 220(2):371--380.
  \doi{10.1023/B:SOLA.0000031405.52911.08}

\bibitem[{{Hagenaar} et~al(2003){Hagenaar}, {Schrijver}, and
  {Title}}]{hagenaar:2003}
{Hagenaar} HJ, {Schrijver} CJ, {Title} AM (2003) {The Properties of Small
  Magnetic Regions on the Solar Surface and the Implications for the Solar
  Dynamo(s)}. \apj 584(2):1107--1119. \doi{10.1086/345792}

\bibitem[{{Hale}(1908)}]{hale:1908}
{Hale} GE (1908) {On the Probable Existence of a Magnetic Field in Sun-Spots}.
  \apj 28:315. \doi{10.1086/141602}

\bibitem[{{Hanasoge}(2022)}]{hanasoge:2022}
{Hanasoge} SM (2022) {Surface and interior meridional circulation in the Sun}.
  Living Reviews in Solar Physics 19(1):3. \doi{10.1007/s41116-022-00034-7}

\bibitem[{{Harvey}(1994)}]{harvey:1994}
{Harvey} KL (1994) {The solar magnetic cycle}. In: {Rutten} RJ, {Schrijver} CJ
  (eds) Solar Surface Magnetism, p 347

\bibitem[{{Harvey-Angle}(1993)}]{harvey:1993}
{Harvey-Angle} KL (1993) "magnetic bipoles on the sun". PhD thesis, -

\bibitem[{{Hathaway}(2012)}]{2012Hathaway}
{Hathaway} DH (2012) {Supergranules as Probes of the Sun's Meridional
  Circulation}. \apj 760(1):84. \doi{10.1088/0004-637X/760/1/84},
  {\href{https://arxiv.org/abs/1210.3343}{{https://arxiv.org/abs/arXiv:1210.3343}}}
  {[astro-ph.SR]}

\bibitem[{{Hathaway} and {Rightmire}(2010)}]{2010HathawayRightmire}
{Hathaway} DH, {Rightmire} L (2010) {Variations in the Sun{\textquoteright}s
  Meridional Flow over a Solar Cycle}. Science 327(5971):1350.
  \doi{10.1126/science.1181990}

\bibitem[{{Hathaway} et~al(2022){Hathaway}, {Upton}, and
  {Mahajan}}]{2022Hathaway_etal}
{Hathaway} DH, {Upton} LA, {Mahajan} SS (2022) {Variations in differential
  rotation and meridional flow within the Sun's surface shear layer
  1996{\textendash}2022}. Frontiers in Astronomy and Space Sciences 9:419.
  \doi{10.3389/fspas.2022.1007290},
  {\href{https://arxiv.org/abs/2212.10619}{{https://arxiv.org/abs/arXiv:2212.10619}}}
  {[astro-ph.SR]}

\bibitem[{{Hayakawa} et~al(2021){Hayakawa}, {Lockwood}, {Owens}, {S{\^o}ma},
  {Besser}, and {van Driel-Gesztelyi}}]{hayakawa_corona_21}
{Hayakawa} H, {Lockwood} M, {Owens} M, et~al (2021) {Graphical evidence for the
  solar coronal structure during the Maunder minimum: comparative study of the
  total eclipse drawings in 1706 and 1715}. J Space Weather Space Climate 11:1.
  \doi{10.1051/swsc/2020035}

\bibitem[{{Hazra} et~al(2023){Hazra}, {Nandy}, {Kitchatinov}, and
  {Choudhuri}}]{hazra:2023}
{Hazra} G, {Nandy} D, {Kitchatinov} L, et~al (2023) {Mean Field Models of Flux
  Transport Dynamo and Meridional Circulation in the Sun and Stars}. \ssr
  219(5):39. \doi{10.1007/s11214-023-00982-y},
  {\href{https://arxiv.org/abs/2302.09390}{{https://arxiv.org/abs/arXiv:2302.09390}}}
  {[astro-ph.SR]}

\bibitem[{{Hindman} et~al(2000){Hindman}, {Haber}, {Toomre}, and
  {Bogart}}]{2000SoPh..192..363H}
{Hindman} B, {Haber} D, {Toomre} J, et~al (2000) {Local Fractional Frequency
  Shifts Used as Tracers of Magnetic Activity}. \solphys 192:363--372.
  \doi{10.1023/A:1005283302728}

\bibitem[{{Hoeksema}(1995)}]{Hoeksema:1995}
{Hoeksema} JT (1995) {The Large-Scale Structure of the Heliospheric Current
  Sheet During the ULYSSES Epoch}. \ssr 72(1-2):137--148.
  \doi{10.1007/BF00768770}

\bibitem[{{Hotta} et~al(2023){Hotta}, {Bekki}, {Gizon}, {Noraz}, and
  {Rast}}]{hotta:2023}
{Hotta} H, {Bekki} Y, {Gizon} L, et~al (2023) {Dynamics of solar large-scale
  flows}. arXiv e-prints arXiv:2307.06481. \doi{10.48550/arXiv.2307.06481},
  {\href{https://arxiv.org/abs/2307.06481}{{https://arxiv.org/abs/arXiv:2307.06481}}}
  {[astro-ph.SR]}

\bibitem[{{Howard} and {LaBonte}(1980)}]{1980ApJ...239L..33H}
{Howard} R, {LaBonte} BJ (1980) {The sun is observed to be a torsional
  oscillator with a period of 11 years}. \apjl 239:L33--L36.
  \doi{10.1086/183286}

\bibitem[{{Howe}(2009)}]{2009LRSP....6....1H}
{Howe} R (2009) {Solar Interior Rotation and its Variation}. Living Reviews in
  Solar Physics 6(1):1. \doi{10.12942/lrsp-2009-1},
  {\href{https://arxiv.org/abs/0902.2406}{{https://arxiv.org/abs/arXiv:0902.2406}}}
  {[astro-ph.SR]}

\bibitem[{{Howe} et~al(2000){Howe}, {Christensen-Dalsgaard}, {Hill}, {Komm},
  {Larsen}, {Schou}, {Thompson}, and {Toomre}}]{2000ApJ...533L.163H}
{Howe} R, {Christensen-Dalsgaard} J, {Hill} F, et~al (2000) {Deeply Penetrating
  Banded Zonal Flows in the Solar Convection Zone}. \apjl 533:L163--L166.
  \doi{10.1086/312623},
  {\href{https://arxiv.org/abs/astro-ph/0003121}{{https://arxiv.org/abs/astro-ph/0003121}}}

\bibitem[{{Howe} et~al(2004){Howe}, {Komm}, {Hill}, {Haber}, and
  {Hindman}}]{2004ApJ...608..562H}
{Howe} R, {Komm} RW, {Hill} F, et~al (2004) {Activity-related Changes in Local
  Solar Acoustic Mode Parameters from Michelson Doppler Imager and Global
  Oscillations Network Group}. \apj 608(1):562--579. \doi{10.1086/392525}

\bibitem[{{Howe} et~al(2009){Howe}, {Christensen-Dalsgaard}, {Hill}, {Komm},
  {Schou}, and {Thompson}}]{2009ApJ...701L..87H}
{Howe} R, {Christensen-Dalsgaard} J, {Hill} F, et~al (2009) {A Note on the
  Torsional Oscillation at Solar Minimum}. \apjl 701(2):L87--L90.
  \doi{10.1088/0004-637X/701/2/L87},
  {\href{https://arxiv.org/abs/0907.2965}{{https://arxiv.org/abs/arXiv:0907.2965}}}
  {[astro-ph.SR]}

\bibitem[{{Howe} et~al(2013){Howe}, {Christensen-Dalsgaard}, {Hill}, {Komm},
  {Larson}, {Rempel}, {Schou}, and {Thompson}}]{2013ApJ...767L..20H}
{Howe} R, {Christensen-Dalsgaard} J, {Hill} F, et~al (2013) {The High-latitude
  Branch of the Solar Torsional Oscillation in the Rising Phase of Cycle 24}.
  \apjl 767(1):L20. \doi{10.1088/2041-8205/767/1/L20}

\bibitem[{{Howe} et~al(2018){Howe}, {Hill}, {Komm}, {Chaplin}, {Elsworth},
  {Davies}, {Schou}, and {Thompson}}]{2018ApJ...862L...5H}
{Howe} R, {Hill} F, {Komm} R, et~al (2018) {Signatures of Solar Cycle 25 in
  Subsurface Zonal Flows}. \apjl 862(1):L5. \doi{10.3847/2041-8213/aad1ed},
  {\href{https://arxiv.org/abs/1807.02398}{{https://arxiv.org/abs/arXiv:1807.02398}}}
  {[astro-ph.SR]}

\bibitem[{Howe et~al(2022)Howe, Chaplin, Christensen-Dalsgaard, Elsworth, and
  Schou}]{Howe_2022}
Howe R, Chaplin WJ, Christensen-Dalsgaard J, et~al (2022) Update on global
  helioseismic observations of the solar torsional oscillation. Research Notes
  of the AAS 6(12):261. \doi{10.3847/2515-5172/aca97d},
  \urlprefix\url{https://dx.doi.org/10.3847/2515-5172/aca97d}

\bibitem[{{Hoyt} and {Schatten}(1998)}]{hoyt98a}
{Hoyt} DV, {Schatten} KH (1998) {Group Sunspot Numbers: A New Solar Activity
  Reconstruction}. Solar Phys 181:491--512. \doi{10.1023/A:1005056326158}

\bibitem[{{Ivanov} and {Miletsky}(2011)}]{ivanov:2011}
{Ivanov} VG, {Miletsky} EV (2011) {Width of Sunspot Generating Zone and
  Reconstruction of Butterfly}. \solphys 268(1):231--242.
  \doi{10.1007/s11207-010-9665-6},
  {\href{https://arxiv.org/abs/1011.4800}{{https://arxiv.org/abs/arXiv:1011.4800}}}
  {[astro-ph.SR]}

\bibitem[{{Jiang} et~al(2009){Jiang}, {Cameron}, {Schmitt}, and
  {Sch{\"u}ssler}}]{Jiang:2009}
{Jiang} J, {Cameron} R, {Schmitt} D, et~al (2009) {Countercell Meridional Flow
  and Latitudinal Distribution of the Solar Polar Magnetic Field}. \apjl
  693(2):L96--L99. \doi{10.1088/0004-637X/693/2/L96}

\bibitem[{{Jiang} et~al(2010){Jiang}, {I{\c{s}}ik}, {Cameron}, {Schmitt}, and
  {Sch{\"u}ssler}}]{jiang:2010}
{Jiang} J, {I{\c{s}}ik} E, {Cameron} RH, et~al (2010) {The Effect of
  Activity-related Meridional Flow Modulation on the Strength of the Solar
  Polar Magnetic Field}. \apj 717(1):597--602.
  \doi{10.1088/0004-637X/717/1/597},
  {\href{https://arxiv.org/abs/1005.5317}{{https://arxiv.org/abs/arXiv:1005.5317}}}
  {[astro-ph.SR]}

\bibitem[{{Jiang} et~al(2014){Jiang}, {Cameron}, and
  {Sch{\"u}ssler}}]{jiang:2014}
{Jiang} J, {Cameron} RH, {Sch{\"u}ssler} M (2014) {Effects of the Scatter in
  Sunspot Group Tilt Angles on the Large-scale Magnetic Field at the Solar
  Surface}. \apj 791(1):5. \doi{10.1088/0004-637X/791/1/5},
  {\href{https://arxiv.org/abs/1406.5564}{{https://arxiv.org/abs/arXiv:1406.5564}}}
  {[astro-ph.SR]}

\bibitem[{{Komm} et~al(2002){Komm}, {Howe}, and {Hill}}]{2002ApJ...572..663K}
{Komm} R, {Howe} R, {Hill} F (2002) {Localizing Width and Energy of Solar
  Global p-Modes}. \apj 572:663--673. \doi{10.1086/340196}

\bibitem[{{Komm} et~al(2018){Komm}, {Howe}, and {Hill}}]{2018SoPh..293..145K}
{Komm} R, {Howe} R, {Hill} F (2018) {Subsurface Zonal and Meridional Flow
  During Cycles 23 and 24}. \solphys 293(10):145.
  \doi{10.1007/s11207-018-1365-7}

\bibitem[{{Krivova} and {Solanki}(2002)}]{krivova:2002}
{Krivova} NA, {Solanki} SK (2002) {The 1.3-year and 156-day periodicities in
  sunspot data: Wavelet analysis suggests a common origin}. \aap 394:701--706.
  \doi{10.1051/0004-6361:20021063}

\bibitem[{{Larmor}(1919)}]{Larmor:1919}
{Larmor} J (1919) {How could a rotating body such as the Sun become magnetic.}
  Rep Brit Assoc Adv Sci 159:160

\bibitem[{{Larson} and {Schou}(2018)}]{2018SoPh..293...29L}
{Larson} TP, {Schou} J (2018) {Global-Mode Analysis of Full-Disk Data from the
  Michelson Doppler Imager and the Helioseismic and Magnetic Imager}. \solphys
  293(2):29. \doi{10.1007/s11207-017-1201-5}

\bibitem[{{Leighton}(1964)}]{Leighton:1964}
{Leighton} RB (1964) {Transport of Magnetic Fields on the Sun.} \apj 140:1547.
  \doi{10.1086/148058}

\bibitem[{{Lekshmi} et~al(2018){Lekshmi}, {Nandy}, and
  {Antia}}]{2018ApJ...861..121L}
{Lekshmi} B, {Nandy} D, {Antia} HM (2018) {Asymmetry in Solar Torsional
  Oscillation and the Sunspot Cycle}. \apj 861(2):121.
  \doi{10.3847/1538-4357/aacbd5},
  {\href{https://arxiv.org/abs/1807.03588}{{https://arxiv.org/abs/arXiv:1807.03588}}}
  {[astro-ph.SR]}

\bibitem[{{Lemerle} and {Charbonneau}(2017)}]{lemerle:2017}
{Lemerle} A, {Charbonneau} P (2017) {A Coupled 2 {\texttimes} 2D
  Babcock-Leighton Solar Dynamo Model. II. Reference Dynamo Solutions}. \apj
  834(2):133. \doi{10.3847/1538-4357/834/2/133},
  {\href{https://arxiv.org/abs/1606.07375}{{https://arxiv.org/abs/arXiv:1606.07375}}}
  {[astro-ph.SR]}

\bibitem[{{Leussu} et~al(2013){Leussu}, {Usoskin}, {Arlt}, and
  {Mursula}}]{leussu13}
{Leussu} R, {Usoskin} IG, {Arlt} R, et~al (2013) {Inconsistency of the Wolf
  sunspot number series around 1848}. Astron Astrophys 559:A28.
  \doi{10.1051/0004-6361/201322373},
  {\href{https://arxiv.org/abs/1310.8443}{{https://arxiv.org/abs/arXiv:1310.8443}}}
  {[astro-ph.SR]}

\bibitem[{{Li}(2018)}]{Li:2018}
{Li} J (2018) {A Systematic Study of Hale and Anti-Hale Sunspot Physical
  Parameters}. \apj 867(2):89. \doi{10.3847/1538-4357/aae31a},
  {\href{https://arxiv.org/abs/1809.08980}{{https://arxiv.org/abs/arXiv:1809.08980}}}
  {[astro-ph.SR]}

\bibitem[{{Libbrecht} and {Woodard}(1990)}]{1990Natur.345..779L}
{Libbrecht} KG, {Woodard} MF (1990) {Solar-cycle effects on solar oscillation
  frequencies}. \nat 345:779--782. \doi{10.1038/345779a0}

\bibitem[{{Maunder}(1903)}]{maunder:1903}
{Maunder} EW (1903) {Spoerer's law of zones}. The Observatory 26:329--330

\bibitem[{{Maunder}(1904)}]{maunder:1904}
{Maunder} EW (1904) {Note on the Distribution of Sun-spots in Heliographic
  Latitude, 1874-1902}. \mnras 64:747--761. \doi{10.1093/mnras/64.8.747}

\bibitem[{{McClintock} and {Norton}(2013)}]{mcclintock:2013}
{McClintock} BH, {Norton} AA (2013) {Recovering Joy's Law as a Function of
  Solar Cycle, Hemisphere, and Longitude}. \solphys 287(1-2):215--227.
  \doi{10.1007/s11207-013-0338-0},
  {\href{https://arxiv.org/abs/1305.3205}{{https://arxiv.org/abs/arXiv:1305.3205}}}
  {[astro-ph.SR]}

\bibitem[{{McClintock} and {Norton}(2016)}]{mcclintock:2016}
{McClintock} BH, {Norton} AA (2016) {Tilt Angle and Footpoint Separation of
  Small and Large Bipolar Sunspot Regions Observed with HMI}. \apj 818(1):7.
  \doi{10.3847/0004-637X/818/1/7},
  {\href{https://arxiv.org/abs/1602.04154}{{https://arxiv.org/abs/arXiv:1602.04154}}}
  {[astro-ph.SR]}

\bibitem[{{McClintock} et~al(2014){McClintock}, {Norton}, and
  {Li}}]{mcclintock:2014}
{McClintock} BH, {Norton} AA, {Li} J (2014) {Re-examining Sunspot Tilt Angle to
  Include Anti-Hale Statistics}. \apj 797(2):130.
  \doi{10.1088/0004-637X/797/2/130},
  {\href{https://arxiv.org/abs/1412.5094}{{https://arxiv.org/abs/arXiv:1412.5094}}}
  {[astro-ph.SR]}

\bibitem[{{Mu{\~n}oz-Jaramillo} et~al(2013){Mu{\~n}oz-Jaramillo},
  {Dasi-Espuig}, {Balmaceda}, and {DeLuca}}]{2013MunozJaramillo_etal}
{Mu{\~n}oz-Jaramillo} A, {Dasi-Espuig} M, {Balmaceda} LA, et~al (2013) {Solar
  Cycle Propagation, Memory, and Prediction: Insights from a Century of
  Magnetic Proxies}. \apjl 767(2):L25. \doi{10.1088/2041-8205/767/2/L25},
  {\href{https://arxiv.org/abs/1304.3151}{{https://arxiv.org/abs/arXiv:1304.3151}}}
  {[astro-ph.SR]}

\bibitem[{{Mu{\~n}oz-Jaramillo} et~al(2021){Mu{\~n}oz-Jaramillo}, {Navarrete},
  and {Campusano}}]{munoz-jaramillo:2021}
{Mu{\~n}oz-Jaramillo} A, {Navarrete} B, {Campusano} LE (2021) {Solar Anti-Hale
  Bipolar Magnetic Regions: A Distinct Population with Systematic Properties}.
  \apj 920(1):31. \doi{10.3847/1538-4357/ac133b},
  {\href{https://arxiv.org/abs/2203.11898}{{https://arxiv.org/abs/arXiv:2203.11898}}}
  {[astro-ph.SR]}

\bibitem[{{Nagy} et~al(2017){Nagy}, {Lemerle}, {Labonville}, {Petrovay}, and
  {Charbonneau}}]{Nagy:2017}
{Nagy} M, {Lemerle} A, {Labonville} F, et~al (2017) {The Effect of ``Rogue''
  Active Regions on the Solar Cycle}. \solphys 292(11):167.
  \doi{10.1007/s11207-017-1194-0},
  {\href{https://arxiv.org/abs/1712.02185}{{https://arxiv.org/abs/arXiv:1712.02185}}}
  {[astro-ph.SR]}

\bibitem[{{Nagy} et~al(2020){Nagy}, {Lemerle}, and {Charbonneau}}]{Nagy:2020}
{Nagy} M, {Lemerle} A, {Charbonneau} P (2020) {Impact of nonlinear surface
  inflows into activity belts on the solar dynamo}. Journal of Space Weather
  and Space Climate 10:62. \doi{10.1051/swsc/2020064}

\bibitem[{{Nelson} et~al(2013){Nelson}, {Brown}, {Brun}, {Miesch}, and
  {Toomre}}]{nelson:2013}
{Nelson} NJ, {Brown} BP, {Brun} AS, et~al (2013) {Magnetic Wreaths and Cycles
  in Convective Dynamos}. \apj 762(2):73. \doi{10.1088/0004-637X/762/2/73},
  {\href{https://arxiv.org/abs/1211.3129}{{https://arxiv.org/abs/arXiv:1211.3129}}}
  {[astro-ph.SR]}

\bibitem[{{Norton} and {Gilman}(2005)}]{norton:2005}
{Norton} AA, {Gilman} PA (2005) {Recovering Solar Toroidal Field Dynamics from
  Sunspot Location Patterns}. \apj 630(2):1194--1205. \doi{10.1086/431961},
  {\href{https://arxiv.org/abs/astro-ph/0506025}{{https://arxiv.org/abs/arXiv:astro-ph/0506025}}}
  {[astro-ph]}

\bibitem[{{Norton} et~al(2014){Norton}, {Charbonneau}, and
  {Passos}}]{norton:2014}
{Norton} AA, {Charbonneau} P, {Passos} D (2014) {Hemispheric Coupling:
  Comparing Dynamo Simulations and Observations}. \ssr 186(1-4):251--283.
  \doi{10.1007/s11214-014-0100-4},
  {\href{https://arxiv.org/abs/1411.7052}{{https://arxiv.org/abs/arXiv:1411.7052}}}
  {[astro-ph.SR]}

\bibitem[{{Norton} et~al(2017){Norton}, {Jones}, {Linton}, and
  {Leake}}]{norton:2017}
{Norton} AA, {Jones} EH, {Linton} MG, et~al (2017) {Magnetic Flux Emergence and
  Decay Rates for Preceder and Follower Sunspots Observed with HMI}. \apj
  842(1):3. \doi{10.3847/1538-4357/aa7052},
  {\href{https://arxiv.org/abs/1705.02053}{{https://arxiv.org/abs/arXiv:1705.02053}}}
  {[astro-ph.SR]}

\bibitem[{{Petrovay}(2010)}]{2010Petrovay}
{Petrovay} K (2010) {Solar Cycle Prediction}. Living Reviews in Solar Physics
  7:6. \doi{10.12942/lrsp-2010-6},
  {\href{https://arxiv.org/abs/1012.5513}{{https://arxiv.org/abs/arXiv:1012.5513}}}
  {[astro-ph.SR]}

\bibitem[{{Petrovay} et~al(2020){Petrovay}, {Nagy}, and
  {Yeates}}]{Petrovay:2020a}
{Petrovay} K, {Nagy} M, {Yeates} AR (2020) {Towards an algebraic method of
  solar cycle prediction. I. Calculating the ultimate dipole contributions of
  individual active regions}. Journal of Space Weather and Space Climate 10:50.
  \doi{10.1051/swsc/2020050},
  {\href{https://arxiv.org/abs/2009.02299}{{https://arxiv.org/abs/arXiv:2009.02299}}}
  {[astro-ph.SR]}

\bibitem[{{Rajaguru} et~al(2001){Rajaguru}, {Basu}, and
  {Antia}}]{2001ApJ...563..410R}
{Rajaguru} SP, {Basu} S, {Antia} HM (2001) {Ring Diagram Analysis of the
  Characteristics of Solar Oscillation Modes in Active Regions}. \apj
  563:410--418. \doi{10.1086/323780},
  {\href{https://arxiv.org/abs/astro-ph/0108227}{{https://arxiv.org/abs/astro-ph/0108227}}}

\bibitem[{{Rempel}(2012)}]{2012ApJ...750L...8R}
{Rempel} M (2012) {High latitude Solar Torsional Oscillations during Phases of
  Changing Magnetic Cycle Amplitude}. \apjl 750(1):L8.
  \doi{10.1088/2041-8205/750/1/L8}

\bibitem[{{Rieger} et~al(1984){Rieger}, {Share}, {Forrest}, {Kanbach},
  {Reppin}, and {Chupp}}]{rieger:1984}
{Rieger} E, {Share} GH, {Forrest} DJ, et~al (1984) {A 154-day periodicity in
  the occurrence of hard solar flares?} \nat 312(5995):623--625.
  \doi{10.1038/312623a0}

\bibitem[{{Rightmire-Upton} et~al(2012){Rightmire-Upton}, {Hathaway}, and
  {Kosak}}]{2012RightmireUpton_etal}
{Rightmire-Upton} L, {Hathaway} DH, {Kosak} K (2012) {Measurements of the Sun's
  High-latitude Meridional Circulation}. \apjl 761(1):L14.
  \doi{10.1088/2041-8205/761/1/L14},
  {\href{https://arxiv.org/abs/1211.0944}{{https://arxiv.org/abs/arXiv:1211.0944}}}
  {[astro-ph.SR]}

\bibitem[{{Schatten} et~al(1978){Schatten}, {Scherrer}, {Svalgaard}, and
  {Wilcox}}]{1978Schatten_atal}
{Schatten} KH, {Scherrer} PH, {Svalgaard} L, et~al (1978) {Using Dynamo Theory
  to predict the sunspot number during Solar Cycle 21}. \grl 5(5):411--414.
  \doi{10.1029/GL005i005p00411}

\bibitem[{{Schou}(1999)}]{1999ApJ...523L.181S}
{Schou} J (1999) {Migration of Zonal Flows Detected Using Michelson Doppler
  Imager F-Mode Frequency Splittings}. \apjl 523:L181--L184.
  \doi{10.1086/312279}

\bibitem[{{Schou} et~al(1998){Schou}, {Antia}, {Basu}, {Bogart}, {Bush},
  {Chitre}, {Christensen-Dalsgaard}, {Di Mauro}, {Dziembowski}, {Eff-Darwich},
  {Gough}, {Haber}, {Hoeksema}, {Howe}, {Korzennik}, {Kosovichev}, {Larsen},
  {Pijpers}, {Scherrer}, {Sekii}, {Tarbell}, {Title}, {Thompson}, and
  {Toomre}}]{1998ApJ...505..390S}
{Schou} J, {Antia} HM, {Basu} S, et~al (1998) {Helioseismic Studies of
  Differential Rotation in the Solar Envelope by the Solar Oscillations
  Investigation Using the Michelson Doppler Imager}. \apj 505(1):390--417.
  \doi{10.1086/306146}

\bibitem[{{Schunker} et~al(2020){Schunker}, {Baumgartner}, {Birch}, {Cameron},
  {Braun}, and {Gizon}}]{schunker:2020}
{Schunker} H, {Baumgartner} C, {Birch} AC, et~al (2020) {Average motion of
  emerging solar active region polarities. II. Joy's law}. \aap 640:A116.
  \doi{10.1051/0004-6361/201937322},
  {\href{https://arxiv.org/abs/2006.05565}{{https://arxiv.org/abs/arXiv:2006.05565}}}
  {[astro-ph.SR]}

\bibitem[{{Schwabe}(1844)}]{schwabe:1844}
{Schwabe} H (1844) {Sonnenbeobachtungen im Jahre 1843. Von Herrn Hofrath
  Schwabe in Dessau}. Astronomische Nachrichten 21(15):233.
  \doi{10.1002/asna.18440211505}

\bibitem[{{Snodgrass} and {Ulrich}(1990)}]{1990ApJ...351..309S}
{Snodgrass} HB, {Ulrich} RK (1990) {Rotation of Doppler Features in the Solar
  Photosphere}. \apj 351:309. \doi{10.1086/168467}

\bibitem[{{Sporer}(1894)}]{sporer:1894}
{Sporer} G (1894) {Number 32. Zehnten Bandes Erstes Stuck. Beobachtungen von
  Sonnenflecken in den Jahren 1885 bis 1893}. Publikationen des
  Astrophysikalischen Observatoriums zu Potsdam 10:3--147

\bibitem[{{Stejko} et~al(2021){Stejko}, {Kosovichev}, and
  {Pipin}}]{stejko:2021}
{Stejko} AM, {Kosovichev} AG, {Pipin} VV (2021) {Forward Modeling Helioseismic
  Signatures of One- and Two-cell Meridional Circulation}. \apj 911(2):90.
  \doi{10.3847/1538-4357/abec70},
  {\href{https://arxiv.org/abs/2101.01220}{{https://arxiv.org/abs/arXiv:2101.01220}}}
  {[astro-ph.SR]}

\bibitem[{{Stenflo}(1970)}]{Stenflo:1970}
{Stenflo} JO (1970) {Hale's Attempts to Determine the Sun's General Magnetic
  Field}. \solphys 14(2):263--273. \doi{10.1007/BF00221312}

\bibitem[{{Stenflo} and {Kosovichev}(2012)}]{stenflo:2012}
{Stenflo} JO, {Kosovichev} AG (2012) {Bipolar Magnetic Regions on the Sun:
  Global Analysis of the SOHO/MDI Data Set}. \apj 745(2):129.
  \doi{10.1088/0004-637X/745/2/129},
  {\href{https://arxiv.org/abs/1112.5226}{{https://arxiv.org/abs/arXiv:1112.5226}}}
  {[astro-ph.SR]}

\bibitem[{{Sun} et~al(2015){Sun}, {Hoeksema}, {Liu}, and {Zhao}}]{sun:2015}
{Sun} X, {Hoeksema} JT, {Liu} Y, et~al (2015) {On Polar Magnetic Field Reversal
  and Surface Flux Transport During Solar Cycle 24}. \apj 798(2):114.
  \doi{10.1088/0004-637X/798/2/114},
  {\href{https://arxiv.org/abs/1410.8867}{{https://arxiv.org/abs/arXiv:1410.8867}}}
  {[astro-ph.SR]}

\bibitem[{{Svalgaard} and {Schatten}(2016)}]{svalgaard16}
{Svalgaard} L, {Schatten} KH (2016) {Reconstruction of the Sunspot Group
  Number: The Backbone Method}. Solar Phys 291:2653--2684.
  \doi{10.1007/s11207-015-0815-8},
  {\href{https://arxiv.org/abs/1506.00755}{{https://arxiv.org/abs/arXiv:1506.00755}}}
  {[astro-ph.SR]}

\bibitem[{{Svalgaard} et~al(1978){Svalgaard}, {Duvall}, and
  {Scherrer}}]{svalgaard:1978}
{Svalgaard} L, {Duvall} JT.~L., {Scherrer} PH (1978) {The strength of the Sun's
  polar fields.} \solphys 58(2):225--239. \doi{10.1007/BF00157268}

\bibitem[{{Svalgaard} et~al(2005){Svalgaard}, {Cliver}, and
  {Kamide}}]{2005Svalgaard_etal}
{Svalgaard} L, {Cliver} EW, {Kamide} Y (2005) {Sunspot cycle 24: Smallest cycle
  in 100 years?} \grl 32(1):L01104. \doi{10.1029/2004GL021664}

\bibitem[{{Thompson} et~al(1996){Thompson}, {Toomre}, {Anderson}, {Antia},
  {Berthomieu}, {Burtonclay}, {Chitre}, {Christensen-Dalsgaard}, {Corbard}, {De
  Rosa}, {Genovese}, {Gough}, {Haber}, {Harvey}, {Hill}, {Howe}, {Korzennik},
  {Kosovichev}, {Leibacher}, {Pijpers}, {Provost}, {Rhodes}, {Schou}, {Sekii},
  {Stark}, and {Wilson}}]{1996Sci...272.1300T}
{Thompson} MJ, {Toomre} J, {Anderson} ER, et~al (1996) {Differential Rotation
  and Dynamics of the Solar Interior}. Science 272(5266):1300--1305.
  \doi{10.1126/science.272.5266.1300}

\bibitem[{{Tlatov} et~al(2013){Tlatov}, {Illarionov}, {Sokoloff}, and
  {Pipin}}]{tlatov:2013}
{Tlatov} A, {Illarionov} E, {Sokoloff} D, et~al (2013) {A new dynamo pattern
  revealed by the tilt angle of bipolar sunspot groups}. \mnras
  432(4):2975--2984. \doi{10.1093/mnras/stt659},
  {\href{https://arxiv.org/abs/1302.2715}{{https://arxiv.org/abs/arXiv:1302.2715}}}
  {[astro-ph.SR]}

\bibitem[{{Tlatova} et~al(2018){Tlatova}, {Tlatov}, {Pevtsov}, {Mursula},
  {Vasil'eva}, {Heikkinen}, {Bertello}, {Pevtsov}, {Virtanen}, and
  {Karachik}}]{tlatova:2018}
{Tlatova} K, {Tlatov} A, {Pevtsov} A, et~al (2018) {Tilt of Sunspot Bipoles in
  Solar Cycles 15 to 24}. \solphys 293(8):118. \doi{10.1007/s11207-018-1337-y},
  {\href{https://arxiv.org/abs/1807.07913}{{https://arxiv.org/abs/arXiv:1807.07913}}}
  {[astro-ph.SR]}

\bibitem[{Ulrich(2010)}]{Ulrich_2010}
Ulrich RK (2010) Solar meridional circulation from doppler shifts of the fe i
  line at 5250 a as measured by the 150-foot solar tower telescope at the mt.
  wilson observatory. The Astrophysical Journal 725(1):658.
  \doi{10.1088/0004-637X/725/1/658},
  \urlprefix\url{https://dx.doi.org/10.1088/0004-637X/725/1/658}

\bibitem[{Ulrich et~al(2022)Ulrich, Tran, and Boyden}]{Ulrich_2022}
Ulrich RK, Tran T, Boyden JE (2022) Polar upwelling at three sunspot minima.
  Research Notes of the {AAS} 6(9):181. \doi{10.3847/2515-5172/ac905f},
  \urlprefix\url{https://doi.org/10.3847/2515-5172/ac905f}

\bibitem[{{Upton} and {Hathaway}(2014{\natexlab{a}})}]{Upton:2014b}
{Upton} L, {Hathaway} DH (2014{\natexlab{a}}) {Effects of Meridional Flow
  Variations on Solar Cycles 23 and 24}. \apj 792(2):142.
  \doi{10.1088/0004-637X/792/2/142},
  {\href{https://arxiv.org/abs/1408.0035}{{https://arxiv.org/abs/arXiv:1408.0035}}}
  {[astro-ph.SR]}

\bibitem[{{Upton} and {Hathaway}(2014{\natexlab{b}})}]{upton:2014a}
{Upton} LA, {Hathaway} DH (2014{\natexlab{b}}) {Predicting the Sun's Polar
  Magnetic Fields with a Surface Flux Transport Model}. \apj 780:5.
  \doi{10.1088/0004-637X/780/1/5},
  {\href{https://arxiv.org/abs/1311.0844}{{https://arxiv.org/abs/arXiv:1311.0844}}}
  {[astro-ph.SR]}

\bibitem[{{Usoskin}(2023)}]{usoskin_LR_23}
{Usoskin} IG (2023) {A History of Solar Activity over Millennia}. Living Rev
  Solar Phys 20:2. \doi{10.1007/s41116-023-00036-z}

\bibitem[{{Usoskin} et~al(2015){Usoskin}, {Arlt}, {Asvestari}, {Hawkins},
  {K{\"a}pyl{\"a}}, {Kovaltsov}, {Krivova}, {Lockwood}, {Mursula}, {O'Reilly},
  {Owens}, {Scott}, {Sokoloff}, {Solanki}, {Soon}, and
  {Vaquero}}]{usoskin_MM_15}
{Usoskin} IG, {Arlt} R, {Asvestari} E, et~al (2015) {The Maunder minimum
  (1645-1715) was indeed a grand minimum: A reassessment of multiple datasets}.
  Astron Astrophys 581:A95. \doi{10.1051/0004-6361/201526652},
  {\href{https://arxiv.org/abs/1507.05191}{{https://arxiv.org/abs/arXiv:1507.05191}}}
  {[astro-ph.SR]}

\bibitem[{{Usoskin} et~al(2016){Usoskin}, {Kovaltsov}, {Lockwood}, {Mursula},
  {Owens}, and {Solanki}}]{usoskin_ADF_16}
{Usoskin} IG, {Kovaltsov} GA, {Lockwood} M, et~al (2016) {A New Calibrated
  Sunspot Group Series Since 1749: Statistics of Active Day Fractions}. Solar
  Phys 291:2685--2708. \doi{10.1007/s11207-015-0838-1},
  {\href{https://arxiv.org/abs/1512.06421}{{https://arxiv.org/abs/arXiv:1512.06421}}}
  {[astro-ph.SR]}

\bibitem[{{van Driel-Gesztelyi} and {Green}(2015)}]{vandriel:2015}
{van Driel-Gesztelyi} L, {Green} LM (2015) {Evolution of Active Regions}.
  Living Reviews in Solar Physics 12(1):1. \doi{10.1007/lrsp-2015-1}

\bibitem[{Vaquero et~al(2016)Vaquero, Svalgaard, Carrasco, Clette, Lef\`evre,
  Gallego, Arlt, Aparicio, Richard, and Howe}]{vaquero16}
Vaquero J, Svalgaard L, Carrasco V, et~al (2016) {A Revised Collection of
  Sunspot Group Numbers}. Solar Phys 291:3061--3074.
  \doi{10.1007/s11207-016-0982-2}

\bibitem[{{Vecchio} et~al(2012){Vecchio}, {Laurenza}, {Meduri}, {Carbone}, and
  {Storini}}]{vecchio:2012}
{Vecchio} A, {Laurenza} M, {Meduri} D, et~al (2012) {The Dynamics of the Solar
  Magnetic Field: Polarity Reversals, Butterfly Diagram, and Quasi-biennial
  Oscillations}. \apj 749(1):27. \doi{10.1088/0004-637X/749/1/27}

\bibitem[{{Verner} et~al(2006){Verner}, {Chaplin}, and
  {Elsworth}}]{2006ApJ...640L..95V}
{Verner} GA, {Chaplin} WJ, {Elsworth} Y (2006) {BiSON Data Show Change in Solar
  Structure with Magnetic Activity}. \apjl 640(1):L95--L98.
  \doi{10.1086/503101}

\bibitem[{Vorontsov et~al(2002)Vorontsov, Christensen-Dalsgaard, Schou,
  Strakhov, and Thompson}]{2002Sci...296..101V}
Vorontsov S, Christensen-Dalsgaard J, Schou J, et~al (2002) Helioseismic
  measurement of solar torsional oscillations. Science 296:101--103.
  \doi{10.1126/science.1069190}

\bibitem[{Waldmeier(1961)}]{waldmeier61}
Waldmeier M (1961) The Sunspot Activity in the Years 1610-1960. Zurich
  Schulthess and Company AG, Z\"urich

\bibitem[{{Wang} and {Sheeley}(1989)}]{wang:1989}
{Wang} YM, {Sheeley} JN.~R. (1989) {Average Properties of Bipolar Magnetic
  Regions during Sunspot CYCLE-21}. \solphys 124(1):81--100.
  \doi{10.1007/BF00146521}

\bibitem[{{Wang} and {Sheeley}(1991)}]{wang:1991}
{Wang} YM, {Sheeley} JN.~R. (1991) {Magnetic Flux Transport and the Sun's
  Dipole Moment: New Twists to the Babcock-Leighton Model}. \apj 375:761.
  \doi{10.1086/170240}

\bibitem[{{Wang} and {Sheeley}(2003)}]{wang:2003}
{Wang} YM, {Sheeley} JN.~R. (2003) {On the Fluctuating Component of the Sun's
  Large-Scale Magnetic Field}. \apj 590(2):1111--1120. \doi{10.1086/375026}

\bibitem[{{Wang} et~al(2015){Wang}, {Colaninno}, {Baranyi}, and
  {Li}}]{wang:2015}
{Wang} YM, {Colaninno} RC, {Baranyi} T, et~al (2015) {Active-region Tilt
  Angles: Magnetic versus White-light Determinations of Joy's Law}. \apj
  798(1):50. \doi{10.1088/0004-637X/798/1/50},
  {\href{https://arxiv.org/abs/1412.2329}{{https://arxiv.org/abs/arXiv:1412.2329}}}
  {[astro-ph.SR]}

\bibitem[{{Watson} and {Basu}(2020)}]{2020ApJ...903L..29W}
{Watson} CB, {Basu} S (2020) {Solar-cycle-related Changes in the Helium
  Ionization Zones of the Sun}. \apjl 903(2):L29.
  \doi{10.3847/2041-8213/abc348},
  {\href{https://arxiv.org/abs/2010.11215}{{https://arxiv.org/abs/arXiv:2010.11215}}}
  {[astro-ph.SR]}

\bibitem[{{Weber} et~al(2011){Weber}, {Fan}, and {Miesch}}]{weber:2011}
{Weber} MA, {Fan} Y, {Miesch} MS (2011) {The Rise of Active Region Flux Tubes
  in the Turbulent Solar Convective Envelope}. \apj 741(1):11.
  \doi{10.1088/0004-637X/741/1/11},
  {\href{https://arxiv.org/abs/1109.0240}{{https://arxiv.org/abs/arXiv:1109.0240}}}
  {[astro-ph.SR]}

\bibitem[{{Weber} et~al(2023){Weber}, {Schunker}, {Jouve}, and
  {I{\c{s}}{\i}k}}]{weber:2023}
{Weber} MA, {Schunker} H, {Jouve} L, et~al (2023) {Understanding Active Region
  Emergence and Origins on the Sun and Other Cool Stars}. arXiv e-prints
  arXiv:2306.06536. \doi{10.48550/arXiv.2306.06536},
  {\href{https://arxiv.org/abs/2306.06536}{{https://arxiv.org/abs/arXiv:2306.06536}}}
  {[astro-ph.SR]}

\bibitem[{{Willamo} et~al(2017){Willamo}, {Usoskin}, and
  {Kovaltsov}}]{willamo17}
{Willamo} T, {Usoskin} IG, {Kovaltsov} GA (2017) {Updated sunspot group number
  reconstruction for 1749-1996 using the active day fraction method}. Astron
  Astrophys 601:A109. \doi{10.1051/0004-6361/201629839},
  {\href{https://arxiv.org/abs/1705.05109}{{https://arxiv.org/abs/arXiv:1705.05109}}}
  {[astro-ph.SR]}

\bibitem[{{Woodard} and {Noyes}(1985)}]{1985Natur.318..449W}
{Woodard} MF, {Noyes} RW (1985) {Change of solar oscillation eigenfrequencies
  with the solar cycle}. \nat 318(6045):449--450. \doi{10.1038/318449a0}

\bibitem[{{Yeates} et~al(2023){Yeates}, {Cheung}, {Jiang}, {Petrovay}, and
  {Wang}}]{yeates:2023}
{Yeates} AR, {Cheung} MCM, {Jiang} J, et~al (2023) {Surface Flux Transport on
  the Sun}. arXiv e-prints arXiv:2303.01209. \doi{10.48550/arXiv.2303.01209},
  {\href{https://arxiv.org/abs/2303.01209}{{https://arxiv.org/abs/arXiv:2303.01209}}}
  {[astro-ph.SR]}

\bibitem[{{Zhao} and {Kosovichev}(2004)}]{zhao:2004}
{Zhao} J, {Kosovichev} AG (2004) {Torsional Oscillation, Meridional Flows, and
  Vorticity Inferred in the Upper Convection Zone of the Sun by Time-Distance
  Helioseismology}. \apj 603(2):776--784. \doi{10.1086/381489}

\bibitem[{{Zhao} et~al(2013){Zhao}, {Bogart}, {Kosovichev}, {Duvall}, and
  {Hartlep}}]{2013Zhao_etal}
{Zhao} J, {Bogart} RS, {Kosovichev} AG, et~al (2013) {Detection of Equatorward
  Meridional Flow and Evidence of Double-cell Meridional Circulation inside the
  Sun}. \apjl 774(2):L29. \doi{10.1088/2041-8205/774/2/L29},
  {\href{https://arxiv.org/abs/1307.8422}{{https://arxiv.org/abs/arXiv:1307.8422}}}
  {[astro-ph.SR]}

\bibitem[{{Zirin}(1988)}]{zirin:1988}
{Zirin} H (1988) {Astrophysics of the sun}. "Cambridge University Press"

\bibitem[{{Zolotova} and {Ponyavin}(2015)}]{zolotova15}
{Zolotova} NV, {Ponyavin} DI (2015) {The Maunder Minimum is Not as Grand as it
  Seemed to be}. Astrophys J 800:42. \doi{10.1088/0004-637X/800/1/42}

\end{thebibliography}
\end{document}